\documentclass[aps,pre,floatfix,twocolumn,superscriptaddress]{revtex4-2}

\usepackage{mathtools}
\usepackage{xcolor}
\usepackage{amsmath}
\usepackage{amsfonts}
\usepackage{amssymb}
\usepackage{graphicx}
\usepackage{parskip}

\begin{document}

\title{Robustness and classical proxy of entanglement in variants of quantum walk}
%\date{}

\author{Christopher Mastandrea}
\affiliation{Department of Physics, University of California, Merced, CA 95343, USA}

\author{Chih-Chun Chien}
\email{cchien5@ucmerced.edu}
\affiliation{Department of Physics, University of California, Merced, CA 95343, USA}

\begin{abstract}
Quantum walk (QW) utilizes its internal quantum states to decide the displacement, thereby introducing single-particle entanglement between the internal and positional degrees of freedom. By simulating three variants of QW with the conventional, symmetric, and split-step translation operators with or without classical randomness in the coin operator, we show the entanglement is robust against both time- and spatially- dependent randomness, which can cause localization transitions of QW. We propose a classical quantity call overlap, which literally measures the overlap between the probability distributions of the internal states, as a proxy of entanglement. The overlap is associated with the off-diagonal terms of the reduced density matrix in the internal space, which then reflects its purity. Therefore, the overlap captures the inverse behavior of the entanglement entropy in most cases. We test the limitation of the classical proxy by constructing a special case with high population imbalance between the internal states to blind the overlap. Possible implications and experimental measurements are also discussed.
\end{abstract}

\maketitle

\section{Introduction}

Quantum walk (QW) extends the classical random walk by incorporating internal degrees of freedom into the walker that determines the spreading of the wave function in real space \cite{QWalkSpringer, Venegas-Andraca2012,10.1145/380752.380757, doi:10.1142/S0219749903000383,KADIAN2021100419,Qiang24}. The dynamics is governed by two quantum operators, the coin and translation operators, and exhibits a characteristic, two-peak probability distribution in contrast to the Gaussian form of classical random walk. Previous works have proposed various uses of the quantum walk, such as providing fine measurements of external magnetic fields \cite{PhysRevA.109.032608}, as an avenue for investigations into topological materials and transport phenomena \cite{PhysRevA.96.033846}, using ballistic spreading speed of QW in quantum algorithms \cite{PhysRevLett.92.120601, PhysRevA.105.042216}, and possible applications in finance~\cite{doi:10.1177/29767032231217444,Chang2024} .

Meanwhile, entanglement is a genuine quantum phenomenon without a faithful classical counterpart. While textbook examples of entanglement usually involve two objects with quantum correlations \cite{MQM,Nielsen_Chuang_2010}, a more general kind of single-particle entanglement (SPE) concerns the entanglement between different degrees of freedom of the same object \cite{single-part-ent-10.1002-qute.202000014, PhysRevA.72.064306, PhysRevA.64.042106}. Moreover, SPE has been shown both through experimental \cite{PhysRevApplied.17.034011} and theoretical studies \cite{PhysRevA.63.012305, PhysRevA.85.022316} to be a promising resource for use within quantum information. Since QW integrates the internal and positional degrees of freedom, previous works \cite{PhysRevLett.111.180503,PhysRevA.99.032320} have found entanglement between those degrees of freedom of the walker.

An interesting path of study in quantum walks is to investigate the influence of classical randomness introduced by imperfections of the apparatus or deliberate controls of the parameters. It has been shown \cite{PhysRevB.96.144204, PhysRevLett.106.180403,PhysRevB.101.144201, PhysRevE.108.024139, PhysRevA.92.052311, PhysRevE.108.024139} that when the classical randomness exceeds a critical value, the QW can exhibit a localization transition characterized by a change of the structures in the probability distributions. When the classical randomness is spatially dependent, the transition is analogous to the Anderson localization in condensed matter systems \cite{PhysRev.109.1492}. There have been works investigating if classical randomness affects the entanglement between the internal and positional degrees of freedom of QW \cite{PhysRevE.108.024139,PhysRevLett.111.180503}. Ref. ~\cite{PhysRevLett.111.180503} studied 1D QW with time-dependent randomness, finding that the dynamic randomness pushes the entanglement entropy~\cite{Nielsen_Chuang_2010}, a typical measure of entanglement, to its maximum faster. The result implies the robustness of quantum entanglement against this type of randomness. In this work, we will systematically investigate the influence of time-dependent and spatially-dependent randomness on various types of QW and show that the robustness is quite common.

Direct measurements of entanglement of quantum systems experimentally, however, have proven to be a difficult task. Some forms of proxy quantities have been used to infer the entanglement, such as statistical methods involving charge fluctuation measurements of non-interacting fermions \cite{PhysRevB.83.161408}, using many-body interference of quantum twins~\cite{Islam2015,Kaufman16}, measuring the Renyi entropy \cite{PhysRevLett.109.020505, PhysRevLett.130.136201}, extracting the eigen-spectrum of the correlation matrix~\cite{Lin2024}, and a potential method using machine learning to identify entanglement within quantum states \cite{khoo_quantum_2021}. These methods typically only work for many-body quantum systems and are not well suited or easily applied to systems with SPE. Moreover, they can be experimentally difficult to implement or computationally expensive for larger systems. Recently, there has been progress in transmitting entangled states through quantum walks with non-Hermitian or topological properties \cite{doi.org-10.100-lpor.202100519, Tang2024, PhysRevResearch.4.043144}. These results show further applications of quantum walks. 

Here, we introduce a classical quantity called overlap, which is literally the overlap between the equal-time probability distributions of the internal states of the walker. We will show that the overlap takes on an inverse relationship with the entanglement entropy of the walker, thereby serving as a classical proxy of the entanglement. We use an explicit construction of the total and reduced density matrices to show that the off-diagonal terms are associated with the overlap, which in turn increases with the purity of the reduced density matrix and determines the entanglement of the total wavefunction. Moreover, we also present an analogy to a composite system of two spin-$1/2$ particles to further explain where and how the entanglement reaches maximum or minimum and why the overlap catches the features. Importantly, the classical nature of the overlap makes it a more 
experimentally accessible quantity that can be used to investigate the entanglement present in QW.
We also present a systematic study of the steady-state entanglement entropy and overlap for a variety of QWs with or without classical randomness of the time- or spatially- dependent type. While the overlap works well for most cases, we deliberately study a special case with high population imbalance of the internal states to show when the overlap may not faithfully track the entanglement.

The rest of the paper is organized as follows. Sec.~\ref{Sec:QWs} summarizes three variants (conventional, symmetric, and split-step) of QW and their typical behavior. Sec.~\ref{Sec:Random} introduces time-dependent and spatially-dependent classical randomness into QW and demonstrates the localization transitions induced by the randomness. Sec.~\ref{Sec:Entanglement} quantifies the entanglement between the internal and positional degrees of freedom by the entanglement entropy and its classical proxy called overlap. Sec.~\ref{Sec:Result} shows the entanglement entropy and overlap of the variants of QW with or without classical randomness. We explain the results by analyzing the total wavefunction of the walker. Sec.~\ref{Sec:Discussion} examines the influence of population imbalance, which may blind the overlap in a special case. Possible implications of experimental realizations and measurements are also discussed. Finally, Sec.~\ref{Sec:Conclusion} concludes our work.

\section{Discrete time quantum walk and its variants}\label{Sec:QWs}
Following Ref.~\cite{QWalkSpringer},
the discrete time quantum walk is governed by two unitary operators, the coin operator $\mathcal{\hat{C}}$ as a rotation operator of the internal states and the translation operator $\mathcal{\hat{T}}$ of the walker. The rotation operator acts to mediate the internal states $|+\rangle = \begin{pmatrix}
    1 \\ 0
\end{pmatrix} $ and $|-\rangle = \begin{pmatrix}
    0 \\ 1
\end{pmatrix} $  at each lattice site. The matrix form is given by $\mathcal{\hat{C}} (\theta, \phi_{1}, \phi_{2}) = \begin{pmatrix} \cos(\theta) & e^{i\phi_{1}}\sin(\theta)  \\ e^{i\phi_{2}} \sin(\theta) & -e^{i(\phi_{1} + \phi_{2}})\cos(\theta) \\ \end{pmatrix}$.
As will be shown shortly, different translation operators give rise to different variants of QW.
We will consider a 1D lattice which has $2N$ total sites ranging from $x=-N, -N+1, \dots, -2, -1, 1, 2, \dots, N-1, N$. 
Open boundary condition of the lattice will be used throughout the paper, and we restrict the time steps so the walker will not exceed the boundary.
The choice is different from the odd-number lattices with a center at $x=0$ used in previous works~\cite{PhysRevA.74.042304, PhysRevA.74.052327,PhysRevE.108.035308}. 
The reason for choosing lattices with even numbers of sites is because we will introduce two variants of QW whose translation operators are only unitary if the lattice does not have a central site at $x=0$ between the left and right halves. 

The wavefunction of the walker is given by $|\psi(t)\rangle =\sum_{x,\sigma}c_{x,\sigma}|x,\sigma\rangle$ with $\sigma=\pm$ and the probability amplitude $c_{x,\sigma}$. 
The initial walker state is a symmetric superposition of the $|\pm\rangle$ states at $x = \pm 1$. Explicitly, $|\psi_{0} \rangle = \sum_{x=\pm 1,\sigma=\pm}\frac{1}{2}|x,\sigma\rangle $. 
We will begin with $\phi_1=\frac{\pi}{2}=\phi_2$, which results in equal populations of the two states during the quantum walk, and in a later discussion we will analyze some cases with other values of $\phi_{1,2}$. The probability distributions at time $t$ for the  $|\pm\rangle $ states can calculated by  $P(x)_{\sigma} = |\langle \sigma|\psi(t)\rangle|^2$, $\sigma=\pm$. The total probability distribution can be found by summing these two distributions, $P(x) = P(x)_{+} + P(x)_{-}$, which should be conserved because only unitary operators are involved.

In the following, we describe three variants of QW that we will investigate in this work. 

\subsection{Conventional quantum walk}
The conventional form as laid out in most discussions of quantum walks has 
the translation operator 
\begin{equation}
\mathcal{\hat{T}} = \sum_{x}(|\psi_{x+1}, +\rangle \langle \psi_{x}, +| + |\psi_{x-1}, -\rangle \langle \psi_{x}, -|).
\end{equation}
This translation operator introduces an asymmetry in how the $|+\rangle$ and $|-\rangle$ states are treated as each state is only ever moved in one direction over the course of the walk. %Because of this asymmetry, it should be noted that we are required to simulate the full lattice of the walk to ensure that all potential interactions are being considered.
The time evolution of the walker follows repeated applications of the translation and rotation operators, so
$|\psi(t+1)\rangle = \mathcal{\hat{T}} \mathcal{\hat{C}} |\psi(t)\rangle$. 
The time-evolved wavefunction of the walker is then given by $|\psi(t)\rangle = \mathcal{\hat{T}} \mathcal{\hat{C}} \mathcal{\hat{T}} \mathcal{\hat{C}} \cdots \mathcal{\hat{T}} \mathcal{\hat{C}} |\psi_{0} \rangle$. 
%= (\mathcal{\hat{T}} \mathcal{\hat{C}})^{N} |\psi_{0} \rangle$. 
For the conventional quantum walk, we only run the walk up to $N-1$ total steps to ensure that the boundary of the lattice will not be traversed.

\begin{figure}

    \centering
    \includegraphics[width=0.85\linewidth]{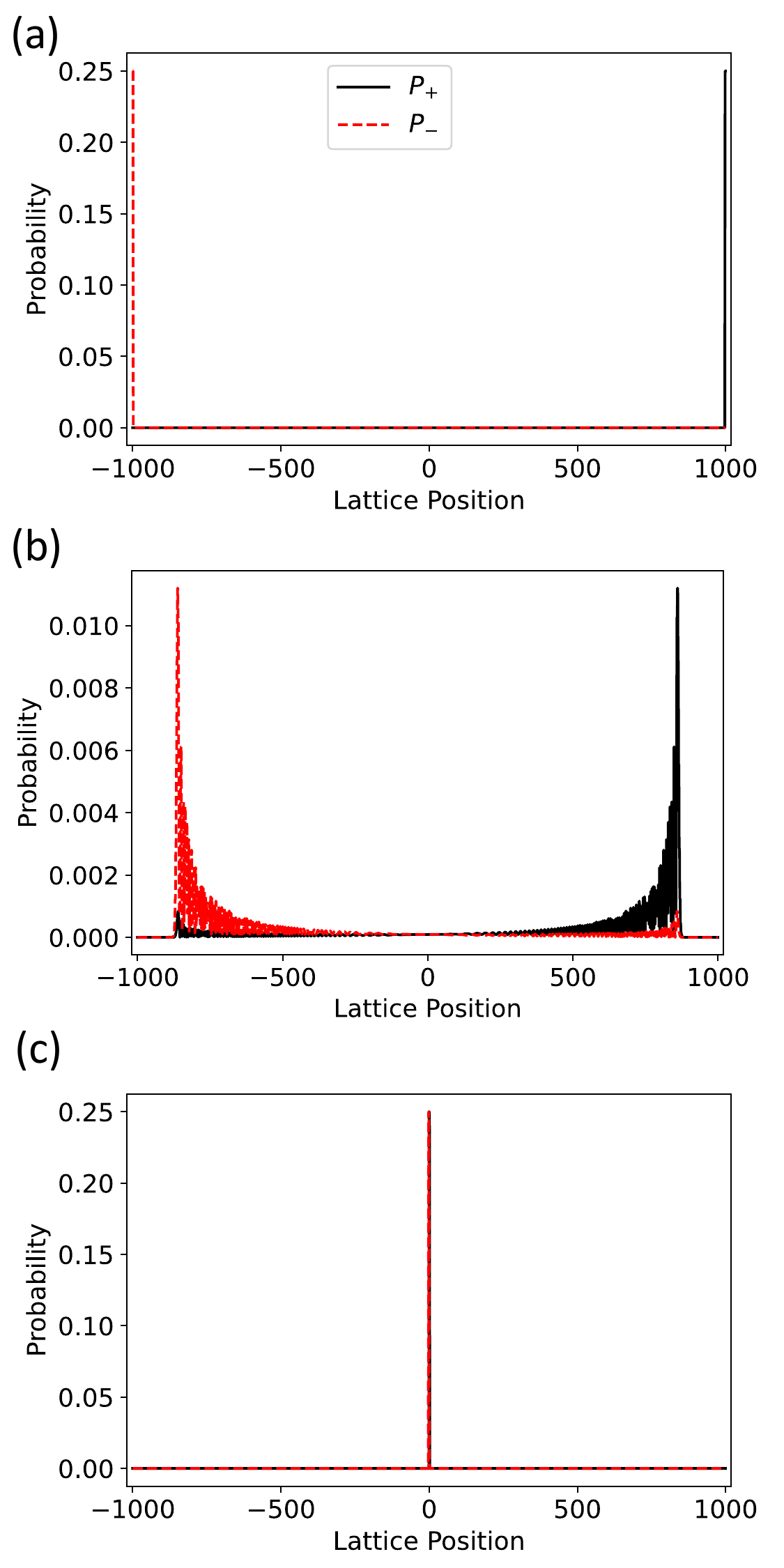}
    \caption{Probability distributions $P_\pm$ of the conventional QW with $N = 1000$ and $\theta = 0$ (a) , $\frac{\pi}{6}$ (b), and $\frac{\pi}{2}$ (c). The black and red spikes coincide in (c). Each spike in (a) and (c) consists of two adjacent peaks.
    }
    \label{fig:conventionalProbs}
    
\end{figure}

Three selected probability distributions from the conventional quantum walk for different values of $\theta$ are shown in Fig. \ref{fig:conventionalProbs}. The distribution in panel (a) shows one of two extreme cases where the walk takes on maximal spreading for $\theta = n\pi$, where $n = 0, 1, 2, \dots$. Panel (c) shows the other extreme case where the walk achieves minimal spreading for $\theta = n\frac{\pi}{2}$, where $n = 1, 2, 3, \dots$. 
In between, the typical behavior of the conventional QW is shown in panel (b) for intermediate values of $\theta$ where the distribution takes on the signature two-peak structure that has been found and studied before \cite{PhysRevA.78.022314, PhysRevLett.92.120601, PhysRevA.105.042216}.

\subsection{Symmetric walk}
Next, we considered a translation operator that does not have any asymmetry in translation imposed by the internal degrees of freedom. 
We first introduce
\begin{equation}
   \mathcal{\hat{T}}_\sigma = \sum_{x>0}|x+1, \sigma \rangle \langle x, \sigma | + \sum_{x<0}|x-1, \sigma \rangle \langle x, \sigma | ,
   \label{eq:translateDef}
\end{equation}
where $\sigma = \pm$.
The symmetric translation operator can then be defined as
\begin{equation}
\mathcal{\hat{T}} = \sum_{\sigma} \mathcal{\hat{T}_{\sigma}}.
\end{equation}
We caution that if the symmetric QW is placed on an odd-number lattice with a center at $x=0$, the translation operator is no longer unitary. 

Similar to the conventional walk, the time-evolved wavefunction of the walker is  $|\psi(t)\rangle = \mathcal{\hat{T}} \mathcal{\hat{C}} \mathcal{\hat{T}} \mathcal{\hat{C}} \cdots \mathcal{\hat{T}} \mathcal{\hat{C}} |\psi_{0} \rangle$. 
%= (\mathcal{\hat{T}} \mathcal{\hat{C}})^{N} |\psi_{0} \rangle$. 
We run the walk on a lattice of $2N$ sites and up to $N-1$ steps to ensure the boundaries are not traversed.
Because of the mirror symmetry of the walk on the lattice, one may simulate only one side of the lattice ($x \geq 1$ or $x \leq 1$) and then duplicate the results to the other side.
%In the case of a $2N$ lattice, this simply means flipping the distribution of the simulated side when copying to the un-simulated side relative to the "origin" of the lattice, i.e. the $N = -1, 1$ locations. 
This trick does not apply to the conventional walk as there is an asymmetry in the distributions ($P_{+}, P_{-}$) on the lattice, as seen in Fig. \ref{fig:conventionalProbs}. 
%Moreover, the symmetric quantum walk is only well-defined on lattices with even numbers of sites. If placed on a lattice with an odd number of sites, one can verify that the translational operator is no longer unitary.

\begin{figure}

    \centering
    \includegraphics[width=0.85\linewidth]{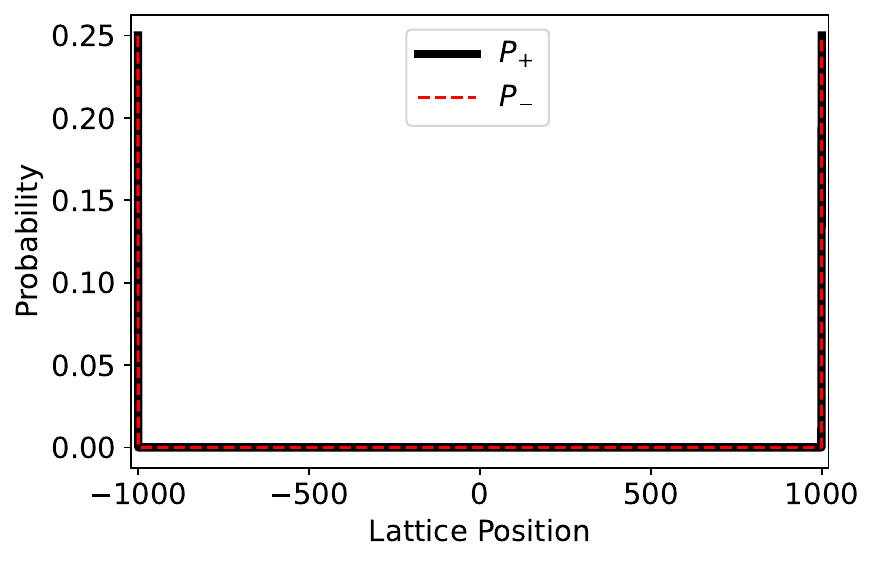}
    \caption{Probability distributions $P_\pm$ of the symmetric QW with $N = 1000$ and $\theta = \frac{\pi}{6}$.}
    \label{fig:SymmFinalProbDists}
    
\end{figure}

The symmetric QW creates probability distributions that resemble two delta functions moving out evenly, as can be seen in Fig.~\ref{fig:SymmFinalProbDists}. This two-peak structure is ubiquitous over different $\theta$ if the initial condition is symmetric about the $|\pm\rangle$ states. This is because $\mathcal{C}(\theta)\left(\begin{array}{c}1/\sqrt{2} \\ 1/\sqrt{2}\end{array}\right)=e^{i\theta}\left(\begin{array}{c}1/\sqrt{2} \\ 1/\sqrt{2}\end{array}\right)$ when $\phi_1=\pi/2=\phi_2$, therefore the two internal states keep the same probabilities as the translation operator shifts their peak locations equally in each step, regardless of the value of $\theta$. The lack of width and asymmetry of the probability distributions is a signature of the symmetric QW, which will result in low entanglement in our later discussion.

\subsection{Split-Step Walk}
The symmetric QW does not have a broadened probability distribution. To modify the behavior, we consider another variant with a split-step translation operator, which has been explored in some previous works \cite{PhysRevA.82.033429, PhysRevA.89.042327, PhysRevA.92.052311}.
The split-step translation operator can be defined as 
\begin{equation}
    \mathcal{\hat{T}}(t) = \begin{cases*}
        \mathcal{\hat{T}}_{+}, & if $t$ is odd, \\
        \mathcal{\hat{T}}_{-},, & if $t$ is even .\\
    \end{cases*}
\end{equation}
Here $\mathcal{\hat{T}}_{\sigma}$ with $\sigma=\pm$ in Eq. \eqref{eq:translateDef} have been used.
Different from the conventional and symmetric QWs, each step of the split-step walk now involves translating all of the $|+\rangle$ states, applying the coin operator, translating the $|-\rangle$ states, and applying the coin operator once again; $|\psi(t+1)\rangle = \mathcal{\hat{C}} \mathcal{\hat{T}}_{-} \mathcal{\hat{C}} \mathcal{\hat{T}}_{+} |\psi(t)\rangle$. One may swap the order of $\pm$ in each time step, and the result is the same as the one without the swap by renaming $\pm$ as $\mp$.
Similar to the symmetric QW, the translation operator of the split-step QW is only unitary if placed on an even-number lattice without a center at $x=0$. When compared to the conventional or symmetric translation operators, the split-step walk shifts the walker by two lattice positions for each step. Therefore, for the split-step quantum walk, we only run the walk up to $\frac{N}{2} - 1$ steps to ensure that we remain inside the boundaries of the lattice.

\begin{figure}

    \includegraphics[width=0.85\linewidth]{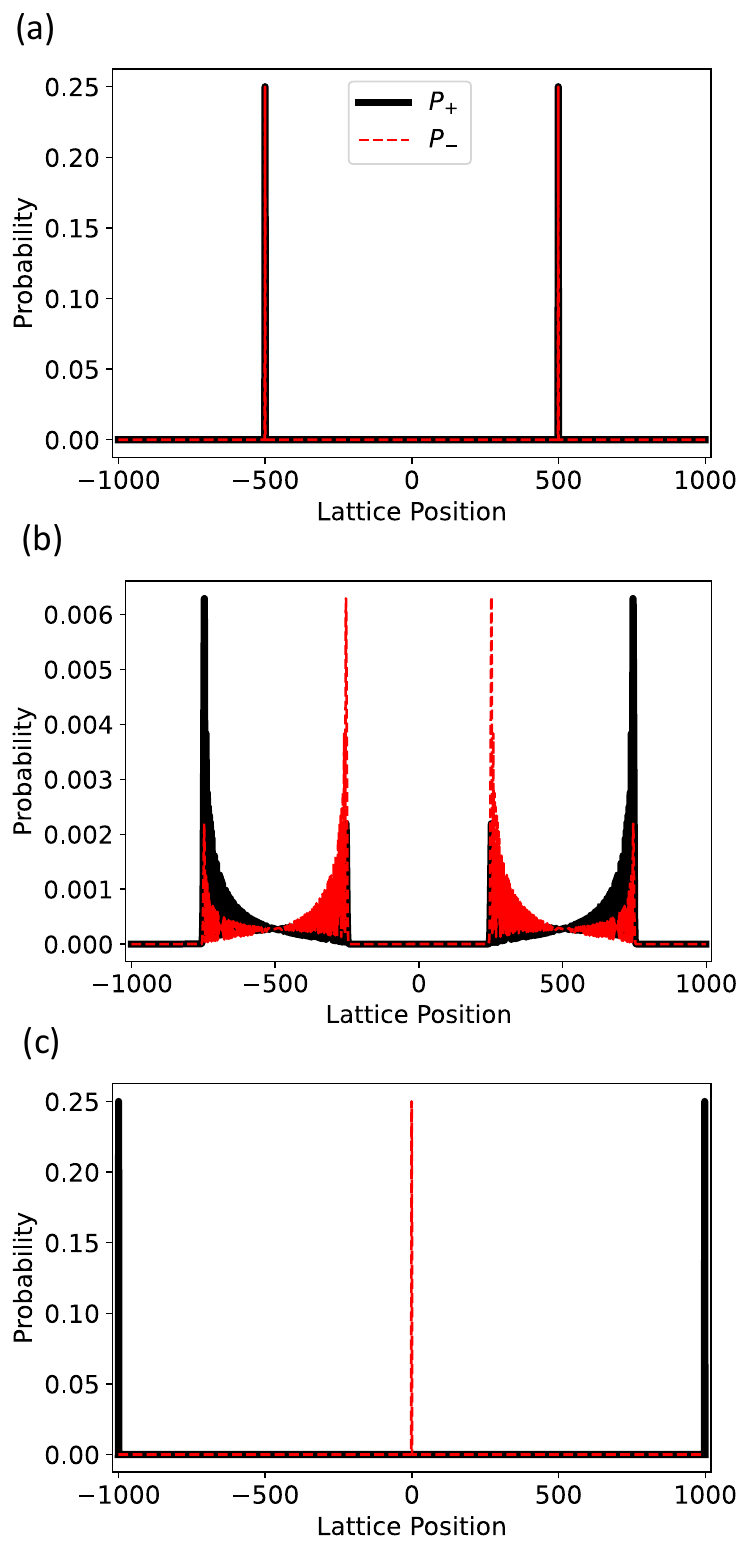}
    \caption{Probability distributions $P_\pm$ of the split-step QW with $N = 1000$ and $\theta = 0$ (a), $\frac{\pi}{6}$ (b), and $\frac{\pi}{2}$ (c). The central spike in (c) contains two distinct, equal and adjacent peaks at $x = \pm 1$.}
    \label{fig:splitStepProbs}
    
\end{figure}

The probability distributions of the split-step walk are shown in Fig. \ref{fig:splitStepProbs} for selected values of $\theta$. Similar to the conventional walk, the split-step walk shows (a) minimal spreading, (b) the traditional two-peak quantum walk structure, and (c) maximal spreading as the value of $\theta$ changes. When compared to the conventional QW, however, we note that the maximal and minimal spreading cases occur at shifted values of $\theta$, with the conventional walk occurring at $\theta = 0$ and $\theta = \frac{\pi}{2}$ for the maximal and minimal cases respectively, and at $\theta = \frac{\pi}{2}$ and $\theta = 0$ for the split-step walk. Furthermore, the split-step walk exhibits double-peak structures on each site of the probability distributions when $0<\theta<\pi/2$ due to its more complicated operations.

% \textbf{write about effect of split-step operator creating two moves each step compared to other walks. leads to needing to run to N/2-1 instead of N-1. Define Step in all walks.}
 
\section{Classical randomness in QW}\label{Sec:Random}
After introducing the three variants of QW, we will add two types of classical randomness to the QWs in order to study their effects on the dynamics and entanglement.

\subsection{Time-dependent random coins}
The first type is a time-dependent random operation on the coin operator by using two rotation operators $\mathcal{\hat{C}}_{1} = \mathcal{C}(\theta_{1}), \mathcal{\hat{C}}_{2} = \mathcal{C}(\theta_{2})$ with rotation angles $\theta_{1} = \theta_{0} + \Delta \theta, \theta_{2} = \theta_{0} - \Delta \theta$, respectively. A classical fair coin is flipped at each step in the walk to decide which rotation operator is applied to the entire lattice at that step in time. Explicitly, this means $P(\mathcal{\hat{C}}_{1}) = P(\mathcal{\hat{C}}_{2}) = \frac{1}{2}$ for every step. In several previous works \cite{PhysRevA.103.032205, PhysRevA.85.012329,ISHAK2021126371}, it has been shown that the two-peak probability distribution of the conventional quantum walk changes to a single-peak distribution when $\Delta\theta$ exceeds a critical value depending on $\theta_0$ and number of steps.

The localization of the conventional QW with time-dependent randomness can be seen in Fig. \ref{fig:localized-distribution} (a), where the probability distributions take the form of Gaussian wave packets when $\Delta\theta$ exceeds the critical value. The distributions of the two internal states, however, do not fully overlap due to the asymmetric translation operators. Meanwhile, the split-step QW in general shows similar behavior as the conventional QW, but the symmetric QW does not localize and maintains the double delta-peak distributions in the presence of time-dependent randomness. As explained in the case without classicial randomness, the symmetry QW does not break the symmetry between the $|\pm\rangle$ states if the initial condition is symmetric about the internal states because the coin operator evenly distributes the internal states for any value of $\theta$ while the translation operator shifts their wavefunctions equally.

\subsection{Spatially dependent random coins}
The second kind of classical randomness that we consider is that of a spatially dependent randomness. In this form, two distinct rotation operators just as those for the time-dependent randomness are used. However, here the coin operator at each site is determined by a fair classical coin to choose from the two rotation angles $\theta_{1}(x), \theta_{2}(x)$ and remain time-independent throughout the walk. In other words, the rotation operator $\mathcal{C}(x)$ is drawn randomly at each site only once in the beginning and used throughout the walk. 
Previous works \cite{PhysRevE.108.024139,PhysRevE.108.035308} have shown that a localization transition occurs when $\Delta\theta$ exceeds a critical value.

\begin{figure}
    \centering
    \includegraphics[width=0.85\linewidth]{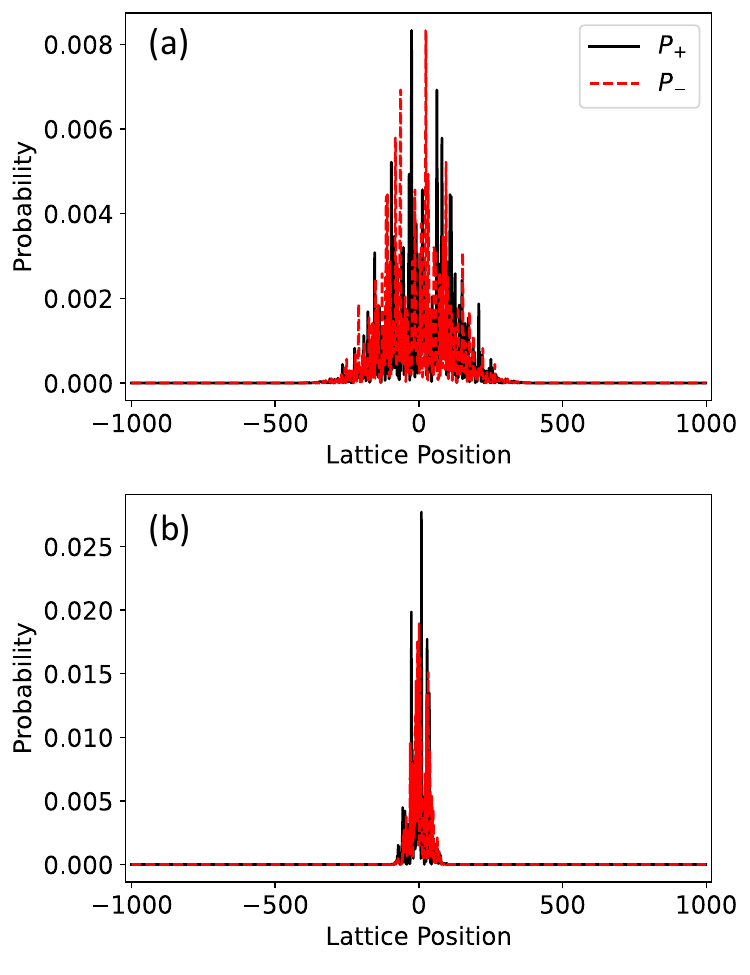}
    \caption{Probability distributions $P_\pm$ of the conventional QW with (a) time-dependent, and (b) spatially-dependent randomness in the coin operator. The two internal states still avoid each other in real space despite the localization. Here $N = 1000$,  $\theta_{0} = \frac{\pi}{6}$, $\Delta \theta = 0.25$ in (a), and $\Delta \theta = 0.3$ in (b).
    }
    \label{fig:localized-distribution}
\end{figure}

The localization of the conventional QW with spatially-dependent randomness can be seen in Fig. \ref{fig:localized-distribution} (b), where the probability distributions take the form of localized exponential wave packets when $\Delta\theta$ exceeds the critical value. Again, the distributions of the two internal states do not fully overlap due to the asymmetric translation operators. The split-step QW in general shows similar behavior as the conventional QW. In contrast, the symmetric QW still exhibits its double delta-peak distributions in the presence of spatially dependent randomness and does not localize. This again is because the coin operator does not break the symmetry between the $|\pm\rangle$ states regardless of $\theta$ if the initial condition is symmetric about the internal states, and the translation operator shifts the wavefunctions equally for both internal states.

In the following discussions which involve classical randomness, we will focus on the conventional and split-step QWs that are well localized in their probability distributions in order to contrast the results with those without the randomness. Importantly, the introduction of classical randomness into quantum walks allows us to test the robustness of the entanglement between the internal and external degrees of freedom of the walker. For example,  Ref.~\cite{PhysRevLett.111.180503} shows the entanglement can survive and reach a maximal value in the presence of time-dependent randomness. We also mention that inhomogeneous coin operators with temporal or spatial patterns but no randomness have been studied in winning strategies of quantum games~\cite{Mittal24}.

%\section{Results and discussions}

\section{Entanglement entropy and its classical proxy}\label{Sec:Entanglement}
The entanglement entropy (ES) between the internal (coin) and external (position) degrees of freedom of the walker can be evaluated as follows. From the total wave function for each step in time written in the form $|\psi_{tot}(x, t)\rangle = \begin{pmatrix} \psi(t)_{+} \\ \psi(t)_{-} \end{pmatrix}$, we construct the density matrix  $\rho_{tot}(t) = |\psi_{tot}(t)\rangle \langle\psi_{tot}(t)|$.
The reduced density matrix of the internal space is found by tracing over the position degree of freedom: $\rho_{s}(t) = Tr_{x}(\rho_{tot}(t))$.
Finally, we calculate the entanglement entropy of the walker by using the  definition~\cite{Nielsen_Chuang_2010}. Explicitly, we evaluate the von Neumann entropy of the reduced density matrix $\rho_{s}$.
\begin{equation}
    \begin{split}
    ES(t) & = -Tr_{s}(\rho_{s} \ln{\rho_{s}}) \\
    &= -\lambda_{+}(t)\ln(\lambda_{+}(t)) - \lambda_{-}(t)\ln(\lambda_{-}(t)),
    \end{split}
    \label{eq:entangleentrop}
\end{equation}
where $\lambda_{\pm}(t)$ are the eigenvalues of the reduced density matrix $\rho_{s}$ for a given step in time.
We remark that we follow a logarithm of base $e$, which differ by a factor of $\approx 0.69$ in the values when compared to other works \cite{ISHAK2021126371,PhysRevA.99.032320, PhysRevA.103.022416} in base $2$.
The typical time evolution of the entanglement entropy of QW can be seen in the upper inset of Fig. \ref{fig:ThreeWalks-noRand}, where the values increases initially but saturates into a steady state. We then evaluate the steady-state value ($\overline{ES}$) by averaging the long-time behavior to ensure the result is free from transient effects.

We remark that in the literature, there have been studies of entanglement by dissecting the lattice into two parts and analyzing the entanglement between the two disjoint parts in real space \cite{PhysRevA.90.023624,PhysRevA.103.022416,Islam2015,Kaufman16}. In contrast, here we focus on single-particle entanglement between the internal and positional degrees of freedom in QW without dissecting the lattice in real space when evaluating the entanglement entropy.

\subsection{Overlap as classical proxy of entanglement}
While the entanglement entropy is a well defined indicator that allows us to investigate and quantify the entanglement present in a given walk, it can be difficult to directly measure experimentally. Here we propose and investigate a classical quantity called overlap that serves as a proxy to the entanglement entropy, allowing for easier experimental verification and measurement of entanglement.

The overlap is defined as
\begin{equation}
    \mathcal{O}(t) = \sum_{x = -N}^{N} P_{+}(x,t) P_{-}(x,t).
    \label{eq:overlap}
\end{equation}
In other words, the above expression measures the overlap of the probability distributions of the two internal states. The reason the overlap reflects the entanglement is as follows. The time-evolved wavefunction of the walker has the form $|\psi (t)\rangle = |\psi_+(x,t)\rangle|+\rangle + |\psi_{-}(x,t)\rangle|-\rangle$, where $|\psi_{\pm}(x,t)\rangle$ represent the wave packets of the $\pm$ components. The total density matrix is then $\rho_{tot}=|\psi (t)\rangle\langle\psi(t)|$. If we use a real-space orthonormal basis $\{ |x\rangle \}$, the reduced density matrix in the internal space is $\rho_s=Tr_{x}(\rho_{tot})=\sum_{x}\langle x|\rho_{tot}|x\rangle$. By using the resolution of the identity, $I=\sum_{x}|x\rangle \langle x|$, the reduced density matrix becomes
\begin{eqnarray}
\rho_s&=&\langle \psi_+ |\psi_+\rangle |+\rangle\langle+| + \langle \psi_- |\psi_+\rangle |-\rangle\langle +|+ \nonumber \\
& &\langle \psi_+ |\psi_-\rangle |+\rangle\langle-|+\langle \psi_- |\psi_-\rangle |-\rangle\langle-|.
\end{eqnarray}
If $|\psi(t)\rangle$ is entangled (separable), $\rho_s$ is a mixed (pure) state with purity $Tr(\rho_s^2)<1$ ($Tr(\rho_s^2)=1$)~\cite{Nielsen_Chuang_2010}.
The entanglement entropy~\eqref{eq:entangleentrop} checks the purity of $\rho_s$ via the von-Neumann entropy formula and reflects the entanglement of $\rho_{tot}$. 
However, a direct calculation shows that the purity of $\rho_s$ is $Tr(\rho_s^2)=|\langle\psi_+|\psi_+\rangle|^2+|\langle\psi_-|\psi_-\rangle|^2+2|\langle\psi_+|\psi_-\rangle|^2$, which increases with the overlap $\mathcal{O}(t)\approx|\langle\psi_+|\psi_-\rangle|^2$. In this perspective, the physical meaning of the overlap~\eqref{eq:overlap} is associated with the magnitude of the off-diagonal terms of $\rho_s$, which reflects the purity of $\rho_s$ in a similar fashion as the entanglement entropy.

Therefore, the maximal (minimal) overlap implies high (low) purity of $\rho_s$, which in turn indicates low (high) entanglement in $|\psi(x,t)\rangle$. For partial overlaps between the two internal states in real space, there is finite entanglement since $\rho_s$ is not completely pure in general. The above argument explains why the overlap serves as a classical proxy to reflect the inverse behavior of the entanglement.
Our simulations of several variants of QW with or without classical randomness will confirm the usefulness of the overlap as a classical proxy for the entanglement entropy in the steady state. As will be shown shortly, the corresponding entanglement entropy is high. However, we caution that the overlap~\eqref{eq:overlap} only involves the measurable densities of the internal states, so there exist situations where it may not faithfully reflect the change in entanglement. We will present a specially designed case to illustrate this subtlety later.

One may also understand the relation between the overlap and entanglement by the resemblance between the state $|\psi (t)\rangle = |\psi_+(x,t)\rangle|+\rangle + |\psi_{-}(x,t)\rangle|-\rangle$ and a composite state of two spin-$1/2$ systems of the form $\frac{1}{\sqrt{2}}(|s_a\rangle_1 |s_b\rangle_2 + |s_c\rangle_1 |s_d\rangle_2)$. Here the two spins are labeled as $1,2$ and $s_j=\pm 1/2$ for $j=a, b, c, d$. The entanglement entropy of the two-spin state will reach maximum for the Bell states and minimum for product states~\cite{Nielsen_Chuang_2010}.
From the analogy between the QW state and two-spin state, the entanglement of QW will be minimum if the two wave packets $|\psi_{\pm}(x,t)\rangle$ have maximal overlap in real space ($|\psi_{\pm}\rangle\approx |\bar{\psi}\rangle$) because the total wavefunction will have the product-state form $|\bar{\psi}\rangle (|+\rangle+|-\rangle)$. In contrast, if the wave packets of the internal states are well separated in real space with minimal overlap, we may treat $|\psi_+\rangle$ as $|"+"\rangle$ and $|\psi_-\rangle$ as $|"-"\rangle$ in real space, so the total wave function has the Bell-state form $(|"+"\rangle|+\rangle + |"-"\rangle|-\rangle)$ with maximal entanglement. Therefore, the overlap indeed serves as an inverse indicator of the entanglement of QW. Our numerical results will confirm the validity of the relation except in some special cases.

The typical time evolution of the overlap of QW is shown in the lower inset of Fig. \ref{fig:ThreeWalks-noRand}. We note that while the overlap does provide an accessible experimental proxy to indicate when a given quantum walk has finite entanglement entropy, it could require a longer time before the overlap settles into a steady state. Moreover, the overlap may have some large spikes within the first few steps of the walk but eventually decays into its steady-state value at longer times. We will show the steady-state value of the overlap $(\overline{\mathcal{O}})$ by taking the average at long times to ensure the result is not influenced by transient values.

\section{Results}\label{Sec:Result}
We begin with $\phi_{1} = \phi_{2} = \frac{\pi}{2}$, which guarantees $P_{+}=P_{-}$ during the walk  with our initial condition for all the cases discussed so far except some QWs with spatially dependent randomness. This can also be seen in Figures \ref{fig:conventionalProbs}, \ref{fig:SymmFinalProbDists}, \ref{fig:splitStepProbs}, and \ref{fig:localized-distribution}(a). In the following, we present the results of entanglement entropy and overlap of those cases.

\subsection{QW without classical randomness}
\begin{figure}
    \centering
    \includegraphics[width=\columnwidth]{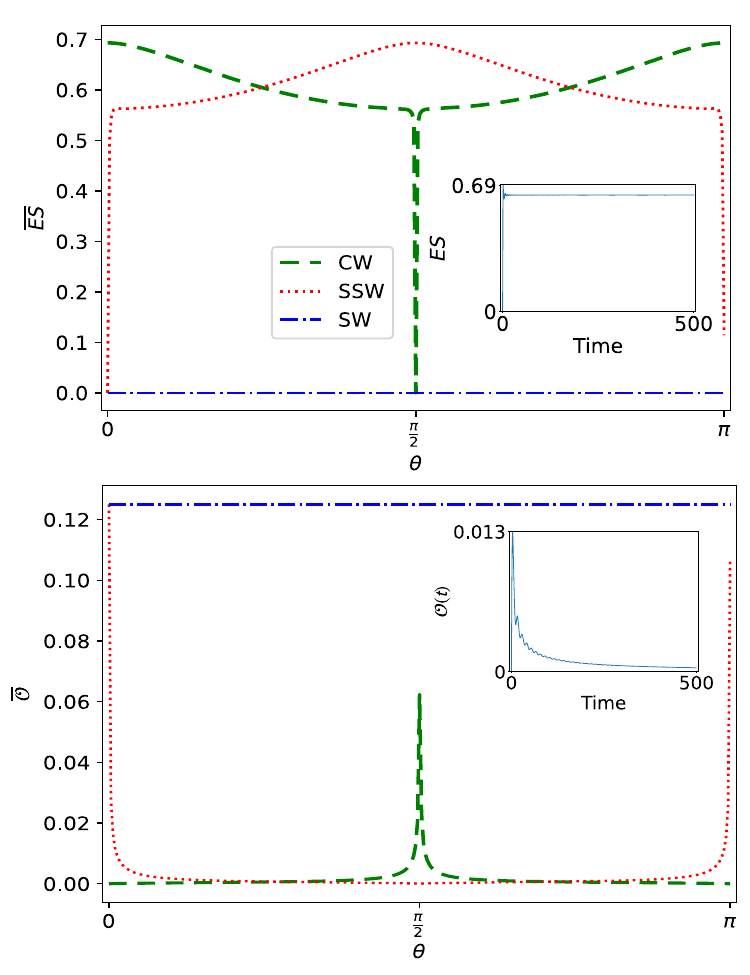}
    \caption{Steady-state entanglement entropy $\overline{ES}$ (top panel) and overlap $\overline{\mathcal{O}}$ (bottom panel) for the three quantum walks without classical randomness. Here $N=500$, and CW, SSW, and SW refer to the conventional walk, split-step walk, and symmetric walk denoted by the dashed, dotted, and dot-dash lines, respectively. The upper (lower) inset shows the behavior of $ES$ ($\mathcal{O}$) as a function of time for the conventional walk with $N=500, \theta=\frac{\pi}{6}$.}
    \label{fig:ThreeWalks-noRand}
\end{figure}

We first investigated the overlap and entanglement entropy for the three QWs without classical randomness. Fig.~\ref{fig:ThreeWalks-noRand} shows the steady-state values of the entanglement entropy and overlap as functions of the parameter $\theta$ of the coin operator. Starting with the symmetric QW, we find that it has vanishing entanglement entropy over all values of $\theta$. Meanwhile, the the overlap quantity is a finite-valued constant, thereby showing an inverse relation to the entanglement entropy. The result can be understood as follows. For the symmetric walk, the spatial (wave packet) part of the wave function is symmetric with respect to the internal states and takes the form $|\psi_+(x,t)\rangle=|\psi_-(x,t)\rangle=|\psi_s(x,t)\rangle$. As a consequence, the total wave function has the form $|\psi(t)\rangle=|\psi_s(x,t)\rangle(|+\rangle+|-\rangle)$, which is a product state. Therefore, the entanglement entropy vanishes while the overlap between the probability distributions of the two internal states reaches maximum.

Next we analyze the conventional QW. We find that the steady-state value of the entanglement entropy (overlap) is maximized (minimized) at $\theta = 0, \pi$ and minimized (maximized) at $\theta = \frac{\pi}{2}$. This can be understood by consulting with the probability distributions shown in Fig.~\ref{fig:conventionalProbs}. When $\theta=0$, the two components are well separated as shown in Fig.~\ref{fig:conventionalProbs}(a). The wave function has the form $|\psi(t)\rangle=|\psi_+(x,t)\rangle|+\rangle+|\psi_-(x,t)\rangle|-\rangle$ with the two wave packets highly concentrated on the opposite ends of the lattice. The wave function thus resembles the Bell state of two entangled spin $1/2$ systems, $|\Psi\rangle=(1/\sqrt{2})(|\uparrow\rangle_1|\uparrow\rangle_2+|\downarrow\rangle_1|\downarrow\rangle_2)$ if we think of the left (right) wave packet as distinct states like  $|\uparrow\rangle_1$ ($|\downarrow\rangle_1$). Therefore, the conventional walk with $\theta=0$ has maximal entanglement but minimal overlap of the probability distributions of the two internal states due to its resemblance of the Bell state. Ref. \cite{PhysRevA.99.032320} investigated the entanglement entropy of a 1D QW and found analytical results in the limit of $t \rightarrow \infty$, which show that the entanglement achieve maximal, steady state values for similar rotation operators to those investigated here.

When $\theta=\pi/2$, the conventional walk does not spread out, as shown in Fig.~\ref{fig:conventionalProbs}(c). The wave packets are then $|\psi_+\rangle=|\psi_-\rangle=|\psi_0\rangle$, where $|\psi_0\rangle$ is the wave packet in the middle of the lattice. The wave function is then $|\psi(t)\rangle=|\psi_0\rangle(|+\rangle+|-\rangle)$, which is of the form of a product state. Therefore, the entanglement entropy vanishes but due to the concentrated wave packets, the overlap reaches maximum in this case. When $0<\theta<\pi/2$, we see from Fig.~\ref{fig:conventionalProbs} (b) that the wave packets of the two internal states spread in opposite directions. As a consequence, the total wavefunction is neither a product state nor a Bell state, so there are both finite entanglement entropy and finite overlap of the probability distributions. Therefore, the behavior of entanglement entropy and overlap shown in Fig.~\ref{fig:ThreeWalks-noRand} can be explained by the analogy to the addition of two spin $1/2$ systems.

From our discussion of the conventional quantum walk, we note that these maxima and minima of entanglement in the steady state correlate with the separation of the probability distributions of the internal states. As a consequence, the overlap correlates inversely with the entanglement entropy and can be used as a classical proxy for entanglement of QW. Furthermore, the overlap also provides a simpler and more experimentally accessible measurement of the elusive entanglement entropy.
 
%We note that the value of entanglement entropy presented in their work differ by ours because of their choice of $\log_{2}$ compared to our choice of $\ln$ as previously discussed.

For the split-step quantum walk, the steady-state value of the entanglement entropy reaches maximum (minimum) when $\theta=\pi/2$ ($\theta=0$), which is opposite to the conventional quantum walk. Meanwhile, the overlap is minimal (maximal) when the entanglement entropy is maximal (minimal), which again confirms the inverse correlation between the two quantities. Interestingly, the entanglement entropy of the split-step QW shows a similar shape as the conventional QW but shifted by $\frac{\pi}{2}$ when plotted as a function of $\theta$. We can use the probability distributions shown in Fig.~\ref{fig:splitStepProbs} to understand the behavior of the entanglement entropy and overlap of the split-step QW. 

When $\theta=0$, the probability distributions shown in Fig.~\ref{fig:splitStepProbs}(a) indicate the wave packets of the two internal states coincide with each other. Therefore, $|\psi_+\rangle=|\psi_-\rangle=|\psi_2\rangle$, where $|\psi_2\rangle$ is the wave packet showing a two-peak structure. The wave function is then $|\psi(t)\rangle=|\psi_2\rangle(|+\rangle+|-\rangle)$, which is a product state with minimal entanglement entropy but maximal overlap of the probability distributions. In contrast, Fig.~\ref{fig:splitStepProbs}(c) indicates that when $\theta=\pi/2$, $|\psi_+\rangle$ has a two-peak structure while $|\psi_-\rangle$ is centered in the middle of the lattice. The total wave function thus has the form $|\psi(t)\rangle=|\psi_+(x,t)\rangle|+\rangle+|\psi_-(x,t)\rangle|-\rangle$ of a Bell state due to the distinct real-space wave packets, thereby its entanglement entropy is maximal but the overlap is minimal. In between $0<\theta<\pi/2$, there is a partial overlap between the two wave packets of the internal states, so there are finite values of the entanglement entropy and overlap. Again, we see that the inverse relation between the entanglement entropy and overlap allows the latter to serve as a classical proxy of entanglement for QW.

% \begin{figure}
%     \centering
%     \includegraphics[width=0.8\linewidth]{Conventional_ES_pi6_N100.pdf}
%     \caption{Entanglement entropy of a conventional walk vs time for $N = 100, \theta=\frac{\pi}{6}$.}
%     \label{fig:ConvetionalESTheta}
% \end{figure}

% \begin{figure}
%     \centering
%     \includegraphics[width=0.8\linewidth]{ESBar_three_walks_P1_N100_pi6.pdf}
%     \caption{Long-time averaged entanglement entropy ($\overline{ES}$) plotted as a function of $\theta$ for three different quantum walks investigated without any classical randomness added. All three walks were preformed with $N = 100$, $\theta = \frac{\pi}{6}$, $\phi_{1} = \phi_{2} = \frac{\pi}{2}$.}
%     \label{fig:enter-label}
% \end{figure}

\subsection{Time-dependent random coins}
\begin{figure}
    \centering
    \includegraphics[width=\columnwidth]{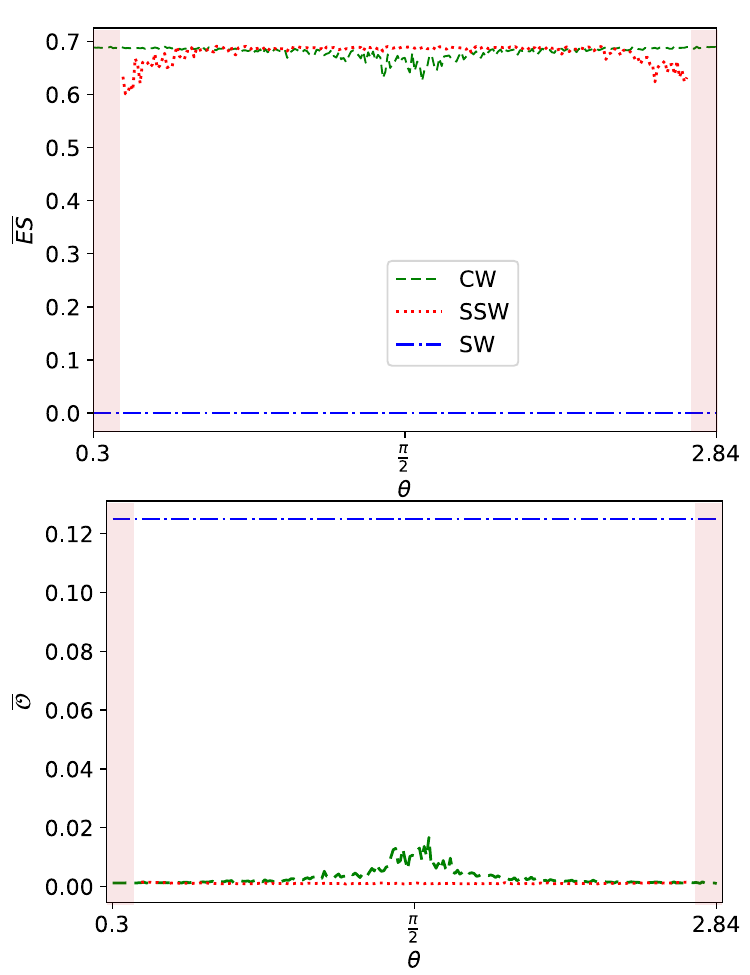}
    \caption{Steady-state entanglement entropy $\overline{ES}$ and overlap $\overline{\mathcal{O}}$ for the three QWs with time-dependent randomness. Here $N=500$ and $\Delta\theta = 0.3$. The shaded regions truncate where the split-step walk does not show localization. CW, SSW, and SW refer to the conventional, split-step, and symmetric walks, respectively.}
    \label{fig:ThreeWalks=time}
\end{figure}

We now introduce classical randomness via time-dependent random rotation operators to the three types of QW.
In Figure~\ref{fig:ThreeWalks=time}, we plot the steady-state values of the entanglement entropy $(\overline{ES})$ and the overlap $(\mathcal{\overline{O}})$ as functions of the rotation operator angle $\theta$ with fixed $\Delta\theta$. For the QWs with classical randomness, however, we focus the systems in the localization regime by choosing a relatively large $\Delta\theta$. Given the rotation angles are randomly chosen from $\theta\pm\Delta\theta$, we only show the range from $\theta = \Delta\theta$ to $\theta = \pi - \Delta\theta$. Further, because we are interested in the entanglement when the walk is fully localized, we discard the results from the split-step QW near $\theta=0$ and $\pi$ where the walk was not fully localized even with the relative large value of $\Delta\theta$.

For the symmetric QW, we find that both the overlap and entanglement entropy are similar to the corresponding case without classical randomness shown in Fig.~\ref{fig:ThreeWalks-noRand}. A close examination shows that the time-dependent random coins only modifies the spreading speed of the symmetric QW while the probability distributions of the internal states remain fully coincide. Therefore, the total wave function still has the form of a product state with zero entanglement and maximal overlap.

In contrast, the conventional QW shows the influence of classical randomness in the form of a relative flatting of the entanglement entropy and the introduction of an amount of fluctuations. Interestingly, the entanglement entropy increases in the presence of time-dependent classical randomness, which is consistent with the observation in Ref.~\cite{PhysRevLett.111.180503}.
Nevertheless, the entanglement entropy shows a similar shape as that without classical randomness and reaches maximal values near $\theta = 0, \pi$ and minimal values around $\theta = \frac{\pi}{2}$. Meanwhile, the overlap exhibits an inverse relationship with the entanglement entropy, showing that even with time-dependent classical randomness and localization, the overlap remains a proxy for the entanglement entropy in the steady state. The reason for the entanglement entropy to survive the time-dependent randomness can be inferred from Fig.~\ref{fig:localized-distribution} (a), which shows that even when the probability distributions become localized, the wave packets of the two internal states still manage to avoid each other locally and keep a minimal overlap. Therefore, the total wave function still has a form similar to the Bell state despite the localized wave packets of both internal states.

Finally, the split-step QW tells a similar story to that of the conventional QW with the randomness appearing to flatten the overall curves for both the entanglement entropy and the overlap. 
%Interestingly, both entanglement entropy and overlap for the split-step QW appear to be more similar to the symmetric QW in the sense that they are nearly constant over the range of $\theta$ values when localized steady states can be obtained. 
The split-step QW produces finite entanglement entropy and minimal overlap because its wave packets of the internal states also manage to avoid each other, making its total wave function more like a Bell state. This reasoning also explains why time-dependent classical randomness does not disrupt the ability for the overlap to act as a classical proxy as it retains the inverse relationship with the entanglement entropy.

% \begin{figure}
%     \centering
%     \includegraphics[width=0.8\linewidth]{ESBar_three_walks_noSymm_N100_pi6.pdf}
%     \caption{\textbf{Not final figure, conventional two-coin is correct, split and symm walk need to be replaced.} Long-time averaged entanglement entropy ($\overline{ES}$) plotted as a function of $\theta$ for three different quantum walks investigated. All walks were preformed with $N = 100$ with the random two-coin walk using $\theta_{0} = \frac{\pi}{6}$. The data for the random two-coin walk was truncated around $\theta = 0, \pi$ as it was difficult to get an estimate of ES due to the large variations present. Similarly, the data around $\theta = \frac{\pi}{2}$ was truncated due to probability distributions not being localized.}
%     \label{fig:AvgESFigure}
% \end{figure}

\subsection{Spatially dependent random coins}
\begin{figure}
    \centering
    \includegraphics[width=\columnwidth]{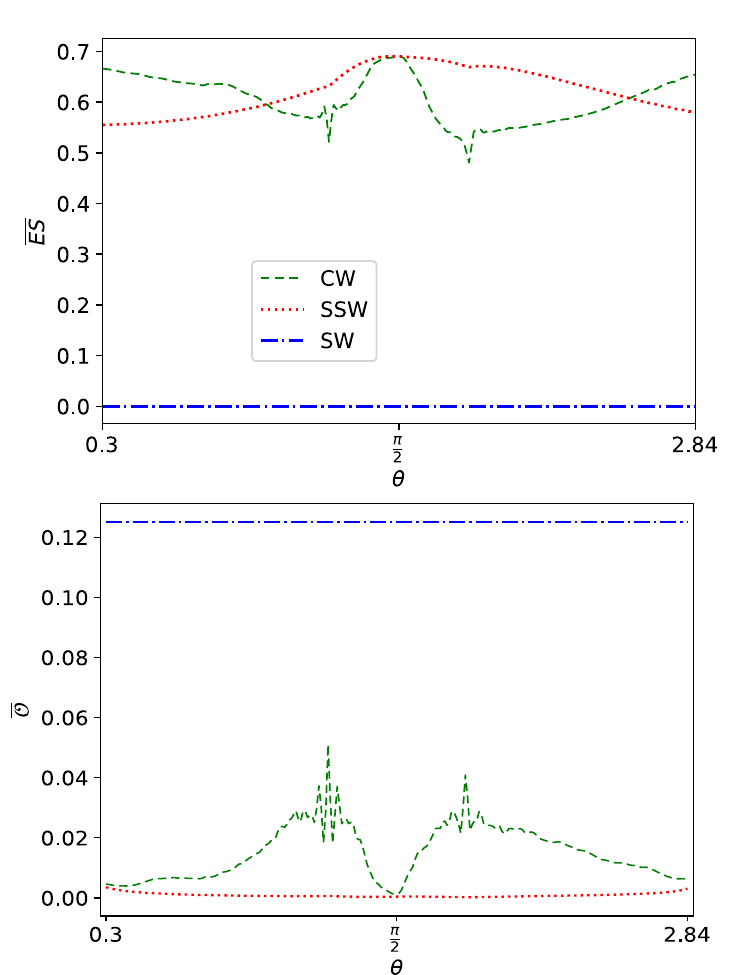}
    \caption{Steady-state entanglement entropy $\overline{ES}$ and overlap $\overline{\mathcal{O}}$ for the three QWs with spatially-dependent randomness. Here $N=500$ and $\Delta\theta = 0.3$. CW, SSW, and SW refer to the conventional walk, split-step walk, and symmetric walk, respectively.}
    \label{fig:ThreeWalks-space}
\end{figure}

Next, we investigate the influence of classical, spatially-dependent randomness on the entanglement entropy and overlap of the three types of QW and show the steady-state values of the entanglement entropy and overlap as functions of $\theta$ with fixed $\Delta\theta$ in Fig.~\ref{fig:ThreeWalks-space}. Once again, the symmetric QW exhibits zero entanglement and finite constant overlap, similar to those observed previously in the no-randomness and time-dependent randomness cases. This is because the probability distributions of the symmetric QW still exhibit the double delta-peak structure with fully overlapped probability distributions of the internal states. Adding time or spatially dependent randomness only changes the locations of the peaks but does not split the distributions of the internal states in real space, which results in the robustness of the symmetric QW against classical randomness.

For the conventional and split-step QWs with spatially dependent randomness, we note that asymmetric patterns with respect to $\theta=\pi/2$ are more prominent, as shown in Fig.~\ref{fig:ThreeWalks-space}. 
%This asymmetry was also found within the time-dependent randomness as can be seen in Fig. \ref{fig:ThreeWalks=time}, however it is not as pronounced of an effect for that case. 
Nevertheless, we find that for both conventional and split-step QWs, the overlap still remains a good proxy for the entanglement entropy and even shows a similar asymmetry that was found within the entanglement entropy. Again the behavior of the entanglement entropy and its negative correlation with the overlap can be understood by writing the total wave function as a sum of the products of the real-space wave packets exemplified in Fig.~\ref{fig:localized-distribution}(b) and the internal states $|\pm\rangle$. A comparison with the product states or Bell states of a composite system then explains where the entanglement entropy (overlap) reaches maximum (minimum). Importantly, the robustness of the entanglement entropy against both time-dependent and spatially-dependent classical randomness has been clearly demonstrated. Moreover, Fig.~\ref{fig:localized-distribution} already indicates low overlaps in the presence of both types of randomness since the two internal states still avoid each other in real space despite the localization. As shown in Figs.~\ref{fig:ThreeWalks=time} and \ref{fig:ThreeWalks-space}, the low overlaps are 
because the wavefunctions are entangled.

\begin{figure}
    \centering
    \includegraphics[width=0.8\linewidth]{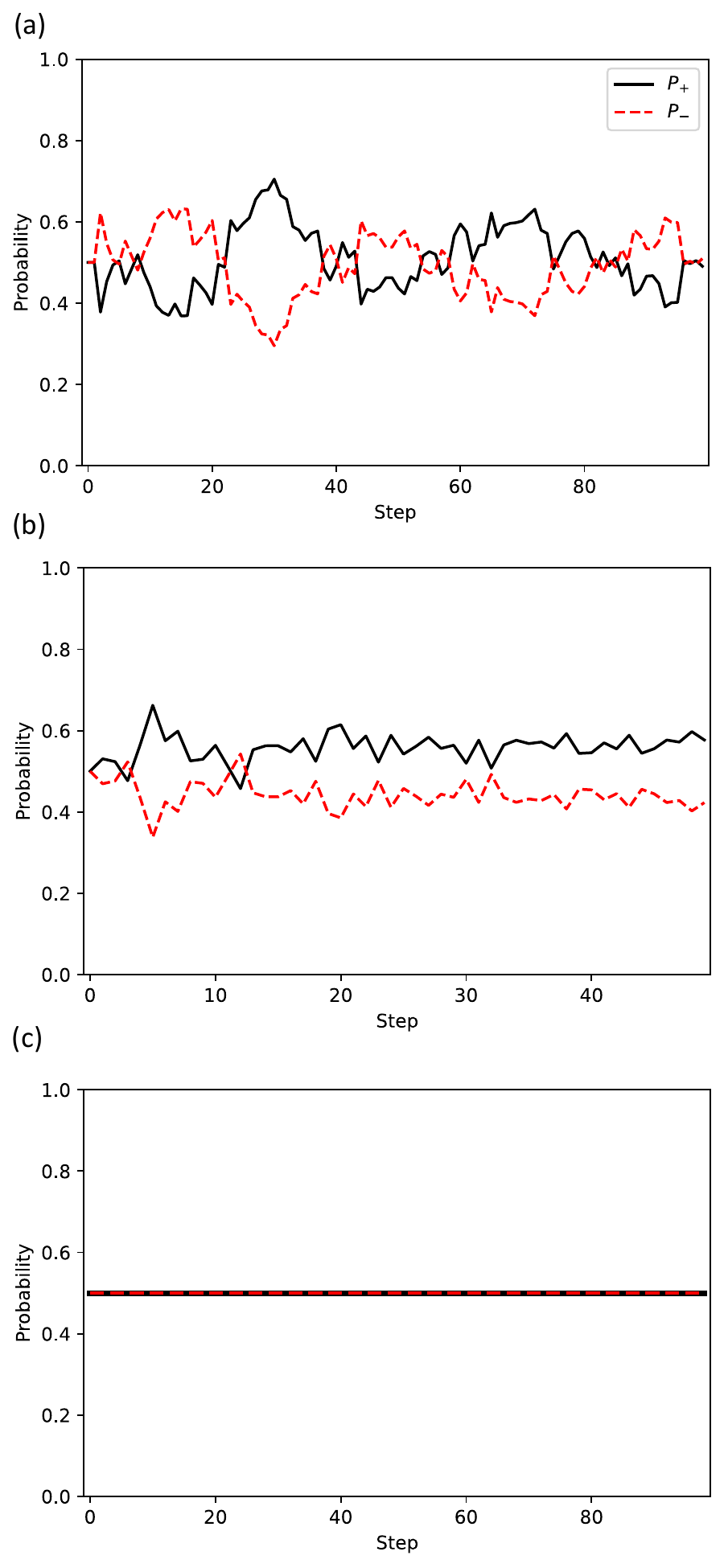}
    \caption{Total probabilities $P_{+}$ (black) and $P_{-}$ (red) at each time step for (a) conventional, (b) split-step, and (c) symmetric QWs with spatially dependent randomness. Here $N=100$,  $\theta = \frac{\pi}{6}$, and $\Delta\theta = 0.3$.
    }
    \label{fig:popimbal-spacerand}
\end{figure}

Fig. \ref{fig:popimbal-spacerand} shows the populations of the two internal states for the three types of QW with spatially dependent randomness. Interestingly, the spatially-dependent randomness may introduce population imbalance of the probability distributions of the internal $|\pm\rangle$ states. While the symmetric QW keeps the populations balanced throughout the walk, the conventional and split-step QWs exhibit population imbalance as time evolves. Nevertheless, the presence of population imbalance does not hinder the overlap to indicate where the entanglement entropy reaches maximum or minimum.

% We note that simply adding in the spatially-dependent randomness is enough to induce an imbalance in the $P_{+}$ and $P_{-}$ probability distributions within the walk.  

\section{Discussion}\label{Sec:Discussion}

% \begin{figure}
%     \centering
%     \includegraphics[width=0.7\linewidth]{Spacerand_splits_popimbal_transFig.pdf}
%     \caption{Total $|+\rangle$ and $|-\rangle$ probabilities, $P_{+}$ and $P_{-}$, at each step in time for a Space-dependent Split Walk ran with $N=500$, $\Delta\theta = $}
%     \label{fig:popimbal-splitstep-generation}
% \end{figure}

\subsection{More about population imbalance}
The possibility of introducing population imbalance in QW via spatially dependent randomness calls for more investigations of this type of behavior. In the following we show that in a special case with high  population imbalance, the overlap may fail to reflect the entanglement entropy, thereby revealing possible limitations of the classical proxy for entanglement.
Without classical randomness, the QW with $\phi_{1} = \phi_{2} = \frac{\pi}{2}$ and a symmetric initial condition keeps the populations balanced during the walk. This can be seen in Figures \ref{fig:conventionalProbs}, \ref{fig:SymmFinalProbDists}, and \ref{fig:splitStepProbs}, where the $P_{+}$ and $P_{-}$ probability distributions are equal to each other. This choice of $\phi_{1} = \phi_{2} = \frac{\pi}{2}$ is a common one that is found in other works because of the balanced populations of the internal states \cite{10.1145/380752.380757, PhysRevA.65.032310}.

\begin{figure}
    \centering
    \includegraphics[width=0.8\linewidth]{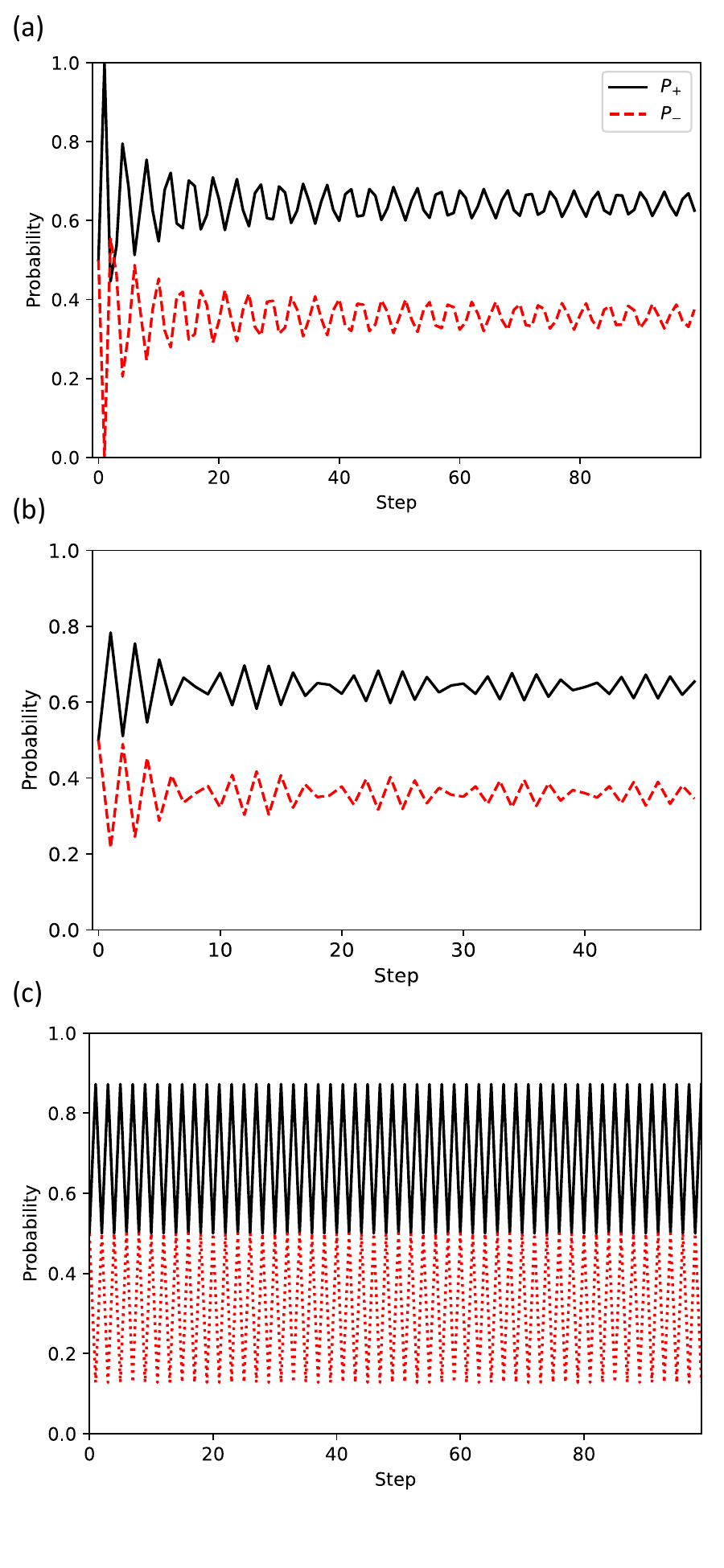}
    \caption{Total probabilities $P_{+}$ (black) and $P_{-}$ (red) at each time step for (a) conventional, (b) split-step, and (c) symmetric QWs without classical randomness. Here $N=100, ~\phi_{1} = \phi_{2} = 0$, and $\theta = 0.84$ (a), $\theta = 2.43$ (b), $\theta = 0.42$ (c).
    }
    \label{fig:popimbal_norand_threepanel}
\end{figure}

Interestingly, population imbalance of QW can be introduced without classical randomness. For example, setting $\phi_{1} = \phi_{2} = 0$ in the coin operator
leads to the evolution of $P_{+}$ and $P_{-}$ shown in Fig.~\ref{fig:popimbal_norand_threepanel} for the three types of QW. we immediately see that this choice introduces population imbalance in all cases without any classical randomness even when the initial condition is symmetric in the two internal states. Importantly, the symmetric QW no longer maintains population balance, in contrast to the previous case shown in Fig.~\ref{fig:popimbal-spacerand}(c). We also remark that for the split-step QW, the majority state depends on the value of $\theta$ and our conventional of which state moves first. The choice of $\phi_{1,2}$ with population imbalance has been used in some works~\cite{JAYAKODY2023100189, Gratsea_2020, PhysRevA.67.052317} as well.

\begin{figure}
    \centering
    \includegraphics[width=\columnwidth]{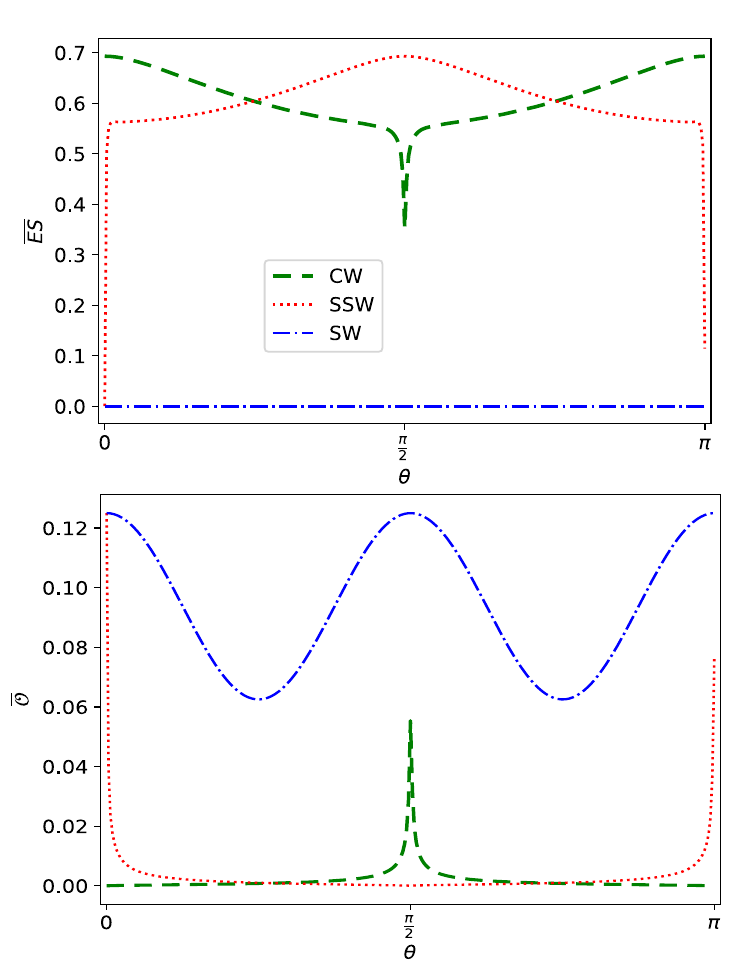}
    \caption{Steady-state entanglement entropy $\overline{ES}$ and overlap $\overline{\mathcal{O}}$ for the three QWs with $\phi_{1} = \phi_{2} = 0$, $N=500$, and no classical randomness. Here CW, SSW, and SW refer to the conventional, split-step, and symmetric QWs, respectively.
    }
    \label{fig:ES-O-popimbal}
\end{figure}

The steady-state entanglement entropy and overlap of the three QWs with population imbalance but no classical randomness are shown in Fig.~\ref{fig:ES-O-popimbal}. 
For the conventional and split-step QWs, the results are nearly identical to those shown before, with the exception of some variation near $\theta = 0, \pi$. The robustness of the entanglement entropy against population imbalance is also evident in the two types of QWs. Moreover, we find that the overlap remains a good proxy for the entanglement entropy, in the sense that the overlap reaches maximum (minimum) when the entanglement entropy reaches minimum (maximum), even in the presence of population imbalance in the conventional and split-step QWs.

\begin{figure}
    \centering
    \includegraphics[width=\columnwidth]{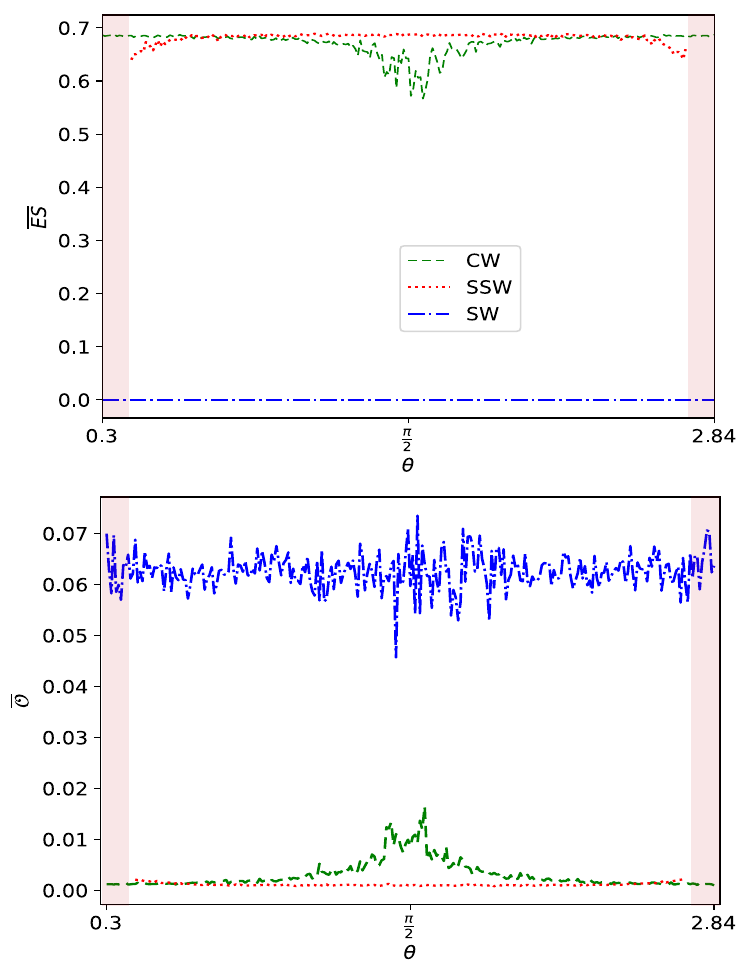}
    \caption{Steady-state entanglement entropy $\overline{ES}$ and overlap $\overline{\mathcal{O}}$ for the three QWs with time-dependent randomness and $\phi_{1} = \phi_{2} = 0$. Like in Figure~\ref{fig:ThreeWalks=time}, the shaded regions truncate where the split-step walk does not show localization. CW, SSW, and SW refer to the conventional, split-step, and symmetric walks, respectively.  Here $N=500$ and $\Delta\theta = 0.3$. }
    \label{fig:ES-O-popimbal-timeRandom}
\end{figure}

\begin{figure}
    \centering
    \includegraphics[width=\columnwidth]{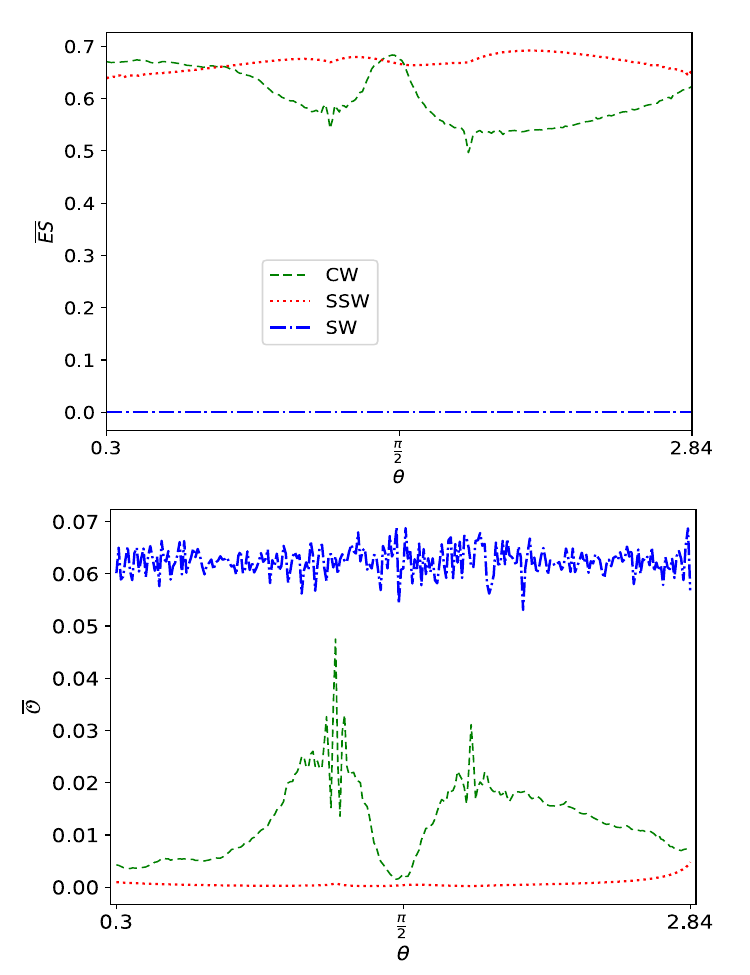}
    \caption{Steady-state entanglement entropy $\overline{ES}$ and overlap $\overline{\mathcal{O}}$ for the three QWs with spatially-dependent randomness and $\phi_{1} = \phi_{2} = 0$. Here $N=500$ and $\Delta\theta = 0.3$. CW, SSW, and SW refer to the conventional, split-step, and symmetric walks, respectively.  
    }
    \label{fig:ES-O-popimbal-spaceRandom}
\end{figure}

% We will consider the three different walks that we investigated before, focusing on the discrete time-dependent versions to see if the overlap working as a proxy for measuring entanglement entropy remains true.

% \textbf{Split-Step walk}
% When considering the split step walk with the time-dependent randomness, we find that entanglement entropy does not produce a steady state with a population imbalance and thus we are not able to effectively calculate the entanglement entropy for this walk. This result shows one of the unfortunate places where it is not possible to find the entanglement entropy of a walk. Moreover, while it is still possible to calculate the overlap of the walk, \textbf{check that the overlap is also essentially useless} it does not reveal any information about the entanglement entropy.  

% \begin{figure}
%     \centering
%     \includegraphics[width=0.9\linewidth]{PopImbal_splitstep_es_theta_0.pdf}
%     \caption{Entanglement Entropy (ES) as a function of time for the Split Step walk with time-dependent randomness. The walk was preformed with $N = 300$, $\theta = \frac{\pi}{6}$, and $\Delta\theta = ??$.}
%     \label{fig:PopImbal_SplitStep_es}
% \end{figure}

% \begin{figure}
%     \centering
%     \includegraphics[width=0.7\linewidth]{PopImbal_symm_overlap_ES_Norandomness.pdf}
%     \caption{Averaged Entanglement entropy $(\overline{ES})$ and Overlap $(\mathcal{O)}$ as a function of $\theta$ for the symmetric walk with no randomness and with $N=100$, $\phi_{1} = \phi_{2} = 0$.}
%     \label{fig:totalSymmESFailure}
% \end{figure}

The symmetric QW shown in Fig.~\ref{fig:ES-O-popimbal}, however, exhibits interesting behavior. One can see that near $\theta=\pi/4, 3\pi/4$, the overlap reaches minimum despite vanishing entanglement entropy. Therefore, the overlap does not reflect the lack of entanglement at those particular values of $\theta$. 
The reason for the failure of the overlap to indicate the entanglement around those special values can be understood by examining the probability distributions shown in Fig.~\ref{fig:popimbal_norand_threepanel}(c). Without the constraint of population balance, the coin operator of the symmetric walk no longer spreads the weights of the internal states evenly when the initial condition is symmetric. Thus, there can be time instances with extremely high population imbalance for some values of $\theta$, where one internal state has nearly zero population. As a consequence, the overlap decreases substantially according to Eq.~\eqref{eq:overlap} in such a situation. Meanwhile, the wave packets of the two internal states still coincide in real space due to the equal shifts by the translation operator of the symmetric walk even in the presence of population imbalance, thereby making the total state more like a product state with low entanglement. 

Therefore, we found a special case of the symmetric QW around particular values of $\theta$ with high population imbalance due to the choice of $\phi_{1,2}$, where the overlap no longer serves as a faithful proxy of the entanglement entropy. This observation demonstrates the limitation of classical proxies, such as the overlap, of genuine quantum objects like the entanglement entropy. Nevertheless, the overlap has been shown to work in the other cases, so one only needs to exercise caution when high population imbalance of the internal states is encountered.

To complete the discussion on QW with population imbalance between the internal states, we show the steady-state entanglement entropy and overlap for the three variants of QW with $\phi_1=0=\phi_2$ and time- or spatially- dependent randomness in the coin operator. Figs. \ref{fig:ES-O-popimbal-timeRandom} and \ref{fig:ES-O-popimbal-spaceRandom} show $\overline{ES}$ and  $\overline{\mathcal{O}}$ in the presence of time-and spatially- dependent randomness with population imbalance, respectively. For both the conventional and the split-step walks, we see a behavior that resembles that which was found in the equal-population cases in Figs. \ref{fig:ThreeWalks-noRand} and \ref{fig:ThreeWalks-space} except the fluctuations due to randomness. 

In contrast, while the entanglement entropy of the symmetric walk resembles that of the equal-population case even in the presence of time- or spatially-dependent randomness, the overlap exhibits fluctuations due to the randomness. We caution that the two-peak probability distribution of the symmetric walk still remains. However, the weights of the two internal states are constantly varying due to the random coin operator when the equal-population condition no longer holds. On the other hand, the noisy behavior renders the highly population-imbalanced configurations negligible in the symmetric walk with time- or spatially-dependent randomness, thereby restoring the overlap as a classical proxy for the entanglement entropy, as shown in Figs.~\ref{fig:ES-O-popimbal-timeRandom} and \ref{fig:ES-O-popimbal-spaceRandom}, even in the presence of population imbalance between the internal states.

% \begin{figure}
%     \centering
%     \includegraphics[width=0.8\linewidth]{PopImbal_symm_overlap_ES.pdf}
%     \caption{Overlap and average entanglement entropy for the symmetric walk. Values were found from a walk with $N=100$, $\Delta\theta = 0.1$.}
%     \label{fig:totalSymmESFailure}
% \end{figure}

\subsection{Implications}
Our analysis of the three variants of QW has showed interesting entanglement between the internal and positional degrees of freedom while the classical proxy by the overlap captures the change of entanglement. Experimental realizations of quantum walks have shown measurements of the probability distributions $P_{\pm}$. For example, Ref. \cite{Su2019} implemented the discrete time quantum walk in a photonic system where the horizontal and vertical photon polarization ($|H\rangle, |V\rangle$) corresponds to the internal $|+ \rangle, |-\rangle$ states, respectively. The probability amplitudes of different polarization can be measured separately, allowing for a reconstruction of the $P_{\pm}$ distributions. There have also been works investigating optical techniques to preform state-dependent imaging of ultracold atoms \cite{PhysRevA.97.023410,PhysRevA.77.033401} and map out the densities of individual internal states. Once the probability distributions $P_\pm$ of the internal states are obtained, the overlap can be evaluated by Eq.~\eqref{eq:overlap}. Moreover, experiments with atomic Bose-Einstein condensates \cite{PhysRevLett.121.070402} have been used to create both quantum walk and classical randomness introduced through the phase factors of the rotation operator. Additionally, Ref. \cite{PhysRevLett.104.050502} studied QW using a photonic setup where the rotation operator is applied using purely passive linear optics while Ref. \cite{IEEE.10141590} constructed QW on solid-state photonic devices.

In addition to simulations of QW by analogue systems using photons and atoms mentioned above, there have been recent attempts and proposals to simulate QW on quantum computers~\cite{PhysRevA.75.062321,Balu_2018,Wadhia2024,Nandi24}. Since currently available quantum computers are mostly of noisy intermediate-scale quantum (NISQ) hardware \cite{Preskill2018quantumcomputingin,RevModPhys.94.015004,AbuGhanem24}, it will be interesting to test the robustness of the probability distributions and entanglement of QW in the presence of randomness on available NISQ hardware.

We have also showed that the overlap is a convenient measure to check single-particle entanglement in future simulations of QW. We emphasize that entanglement is a genuine quantum property~\cite{MQM,Nielsen_Chuang_2010} and therefore cannot be faithfully represented by a classical quantity like the overlap under all circumstances. Nevertheless, our analysis of the entanglement entropy and overlap through the total wavefunction from the wave packets in real space and internal states of the walker shows that the classical proxy in general reflects the entanglement of QW. Unless the system exhibits high population imbalance which may blind the overlap due to its construction, the measurement of entanglement of QW via the classical proxy can be relatively straightforward when compared to other fully quantum measures.

\section{Conclusion}\label{Sec:Conclusion}
The simulations of the three variants (conventional, symmetric, and split-step) of QW have demonstrated that single-particle entanglement between the internal and positional degrees of freedom is robust against different types of classical randomness in both delocalized and localized regimes.
The classical quantity called overlap acts as a proxy for the entanglement entropy of the walker, allowing for additional experimentally accessible and achievable methods of quantifying entanglement. Even though the overlap captures the behavior of entanglement in most situations, the genuine quantum nature of the entanglement is reflected by a special case with high population imbalance between the internal states which blinds the overlap. We envision advancements in characterization and manipulation of single-particle entanglement utilizing QW will find more applications in future quantum technologies.

\begin{acknowledgments}
This work was supported
by the National Science Foundation under Grant No.
PHY-2310656.
\end{acknowledgments}

%\bibliographystyle{apsrev}
%\bibliography{Reference}

\begin{thebibliography}{67}%
	\makeatletter
	\providecommand \@ifxundefined [1]{%
		\@ifx{#1\undefined}
	}%
	\providecommand \@ifnum [1]{%
		\ifnum #1\expandafter \@firstoftwo
		\else \expandafter \@secondoftwo
		\fi
	}%
	\providecommand \@ifx [1]{%
		\ifx #1\expandafter \@firstoftwo
		\else \expandafter \@secondoftwo
		\fi
	}%
	\providecommand \natexlab [1]{#1}%
	\providecommand \enquote  [1]{``#1''}%
	\providecommand \bibnamefont  [1]{#1}%
	\providecommand \bibfnamefont [1]{#1}%
	\providecommand \citenamefont [1]{#1}%
	\providecommand \href@noop [0]{\@secondoftwo}%
	\providecommand \href [0]{\begingroup \@sanitize@url \@href}%
	\providecommand \@href[1]{\@@startlink{#1}\@@href}%
	\providecommand \@@href[1]{\endgroup#1\@@endlink}%
	\providecommand \@sanitize@url [0]{\catcode `\\12\catcode `\$12\catcode
		`\&12\catcode `\#12\catcode `\^12\catcode `\_12\catcode `\%12\relax}%
	\providecommand \@@startlink[1]{}%
	\providecommand \@@endlink[0]{}%
	\providecommand \url  [0]{\begingroup\@sanitize@url \@url }%
	\providecommand \@url [1]{\endgroup\@href {#1}{\urlprefix }}%
	\providecommand \urlprefix  [0]{URL }%
	\providecommand \Eprint [0]{\href }%
	\providecommand \doibase [0]{https://doi.org/}%
	\providecommand \selectlanguage [0]{\@gobble}%
	\providecommand \bibinfo  [0]{\@secondoftwo}%
	\providecommand \bibfield  [0]{\@secondoftwo}%
	\providecommand \translation [1]{[#1]}%
	\providecommand \BibitemOpen [0]{}%
	\providecommand \bibitemStop [0]{}%
	\providecommand \bibitemNoStop [0]{.\EOS\space}%
	\providecommand \EOS [0]{\spacefactor3000\relax}%
	\providecommand \BibitemShut  [1]{\csname bibitem#1\endcsname}%
	\let\auto@bib@innerbib\@empty
	%</preamble>
	\bibitem [{\citenamefont {Wang}\ and\ \citenamefont
		{Manouchehri}(2014)}]{QWalkSpringer}%
	\BibitemOpen
	\bibfield  {author} {\bibinfo {author} {\bibfnamefont {J.}~\bibnamefont
			{Wang}}\ and\ \bibinfo {author} {\bibfnamefont {K.}~\bibnamefont
			{Manouchehri}},\ }\href@noop {} {\emph {\bibinfo {title} {Physical
				Implementations of Quantum Walks}}}\ (\bibinfo  {publisher} {Springer},\
	\bibinfo {address} {Berlin, Germany},\ \bibinfo {year} {2014})\BibitemShut
	{NoStop}%
	\bibitem [{\citenamefont {Venegas-Andraca}(2012)}]{Venegas-Andraca2012}%
	\BibitemOpen
	\bibfield  {author} {\bibinfo {author} {\bibfnamefont {S.~E.}\ \bibnamefont
			{Venegas-Andraca}},\ }\bibfield  {title} {\bibinfo {title} {Quantum walks: a
			comprehensive review},\ }\href {https://doi.org/10.1007/s11128-012-0432-5}
	{\bibfield  {journal} {\bibinfo  {journal} {Quantum Inf. Process.}\ }\textbf
		{\bibinfo {volume} {11}},\ \bibinfo {pages} {1015} (\bibinfo {year}
		{2012})}\BibitemShut {NoStop}%
	\bibitem [{\citenamefont {Ambainis}\ \emph {et~al.}(2001)\citenamefont
		{Ambainis}, \citenamefont {Bach}, \citenamefont {Nayak}, \citenamefont
		{Vishwanath},\ and\ \citenamefont {Watrous}}]{10.1145/380752.380757}%
	\BibitemOpen
	\bibfield  {author} {\bibinfo {author} {\bibfnamefont {A.}~\bibnamefont
			{Ambainis}}, \bibinfo {author} {\bibfnamefont {E.}~\bibnamefont {Bach}},
		\bibinfo {author} {\bibfnamefont {A.}~\bibnamefont {Nayak}}, \bibinfo
		{author} {\bibfnamefont {A.}~\bibnamefont {Vishwanath}},\ and\ \bibinfo
		{author} {\bibfnamefont {J.}~\bibnamefont {Watrous}},\ }\bibfield  {title}
	{\bibinfo {title} {One-dimensional quantum walks},\ }in\ \href
	{https://doi.org/10.1145/380752.380757} {\emph {\bibinfo {booktitle}
			{Proceedings of the Thirty-Third Annual ACM Symposium on Theory of
				Computing}}},\ \bibinfo {series and number} {STOC '01}\ (\bibinfo
	{publisher} {Association for Computing Machinery},\ \bibinfo {address} {New
		York, NY, USA},\ \bibinfo {year} {2001})\ p.\ \bibinfo {pages}
	{37–49}\BibitemShut {NoStop}%
	\bibitem [{\citenamefont {Ambainis}(2003)}]{doi:10.1142/S0219749903000383}%
	\BibitemOpen
	\bibfield  {author} {\bibinfo {author} {\bibfnamefont {A.}~\bibnamefont
			{Ambainis}},\ }\bibfield  {title} {\bibinfo {title} {Quantum walks and their
			algorithmic applications},\ }\href
	{https://doi.org/10.1142/S0219749903000383} {\bibfield  {journal} {\bibinfo
			{journal} {Int. J. Quantum Inf.}\ }\textbf {\bibinfo {volume} {01}},\
		\bibinfo {pages} {507} (\bibinfo {year} {2003})},\ \Eprint
	{https://arxiv.org/abs/https://doi.org/10.1142/S0219749903000383}
	{https://doi.org/10.1142/S0219749903000383} \BibitemShut {NoStop}%
	\bibitem [{\citenamefont {Kadian}\ \emph {et~al.}(2021)\citenamefont {Kadian},
		\citenamefont {Garhwal},\ and\ \citenamefont {Kumar}}]{KADIAN2021100419}%
	\BibitemOpen
	\bibfield  {author} {\bibinfo {author} {\bibfnamefont {K.}~\bibnamefont
			{Kadian}}, \bibinfo {author} {\bibfnamefont {S.}~\bibnamefont {Garhwal}},\
		and\ \bibinfo {author} {\bibfnamefont {A.}~\bibnamefont {Kumar}},\ }\bibfield
	{title} {\bibinfo {title} {Quantum walk and its application domains: A
			systematic review},\ }\href
	{https://doi.org/https://doi.org/10.1016/j.cosrev.2021.100419} {\bibfield
		{journal} {\bibinfo  {journal} {Comput. Sci. Rev}\ }\textbf {\bibinfo
			{volume} {41}},\ \bibinfo {pages} {100419} (\bibinfo {year}
		{2021})}\BibitemShut {NoStop}%
	\bibitem [{\citenamefont {Qiang}\ \emph {et~al.}(2024)\citenamefont {Qiang},
		\citenamefont {Ma},\ and\ \citenamefont {Song}}]{Qiang24}%
	\BibitemOpen
	\bibfield  {author} {\bibinfo {author} {\bibfnamefont {X.}~\bibnamefont
			{Qiang}}, \bibinfo {author} {\bibfnamefont {S.}~\bibnamefont {Ma}},\ and\
		\bibinfo {author} {\bibfnamefont {H.}~\bibnamefont {Song}},\ }\href@noop {}
	{\bibinfo {title} {Review on quantum walk computing: Theory, implementation,
			and application}} (\bibinfo {year} {2024}),\ \bibinfo {note} {arXiv:
		2404.04178}\BibitemShut {NoStop}%
	\bibitem [{\citenamefont {Shukla}\ and\ \citenamefont
		{Chandrashekar}(2024)}]{PhysRevA.109.032608}%
	\BibitemOpen
	\bibfield  {author} {\bibinfo {author} {\bibfnamefont {K.}~\bibnamefont
			{Shukla}}\ and\ \bibinfo {author} {\bibfnamefont {C.~M.}\ \bibnamefont
			{Chandrashekar}},\ }\bibfield  {title} {\bibinfo {title} {Quantum
			magnetometry using discrete-time quantum walk},\ }\href
	{https://doi.org/10.1103/PhysRevA.109.032608} {\bibfield  {journal} {\bibinfo
			{journal} {Phys. Rev. A}\ }\textbf {\bibinfo {volume} {109}},\ \bibinfo
		{pages} {032608} (\bibinfo {year} {2024})}\BibitemShut {NoStop}%
	\bibitem [{\citenamefont {Barkhofen}\ \emph {et~al.}(2017)\citenamefont
		{Barkhofen}, \citenamefont {Nitsche}, \citenamefont {Elster}, \citenamefont
		{Lorz}, \citenamefont {G\'abris}, \citenamefont {Jex},\ and\ \citenamefont
		{Silberhorn}}]{PhysRevA.96.033846}%
	\BibitemOpen
	\bibfield  {author} {\bibinfo {author} {\bibfnamefont {S.}~\bibnamefont
			{Barkhofen}}, \bibinfo {author} {\bibfnamefont {T.}~\bibnamefont {Nitsche}},
		\bibinfo {author} {\bibfnamefont {F.}~\bibnamefont {Elster}}, \bibinfo
		{author} {\bibfnamefont {L.}~\bibnamefont {Lorz}}, \bibinfo {author}
		{\bibfnamefont {A.}~\bibnamefont {G\'abris}}, \bibinfo {author}
		{\bibfnamefont {I.}~\bibnamefont {Jex}},\ and\ \bibinfo {author}
		{\bibfnamefont {C.}~\bibnamefont {Silberhorn}},\ }\bibfield  {title}
	{\bibinfo {title} {Measuring topological invariants in disordered
			discrete-time quantum walks},\ }\href
	{https://doi.org/10.1103/PhysRevA.96.033846} {\bibfield  {journal} {\bibinfo
			{journal} {Phys. Rev. A}\ }\textbf {\bibinfo {volume} {96}},\ \bibinfo
		{pages} {033846} (\bibinfo {year} {2017})}\BibitemShut {NoStop}%
	\bibitem [{\citenamefont {Blanchard}\ and\ \citenamefont
		{Hongler}(2004)}]{PhysRevLett.92.120601}%
	\BibitemOpen
	\bibfield  {author} {\bibinfo {author} {\bibfnamefont {P.}~\bibnamefont
			{Blanchard}}\ and\ \bibinfo {author} {\bibfnamefont {M.-O.}\ \bibnamefont
			{Hongler}},\ }\bibfield  {title} {\bibinfo {title} {Quantum random walks and
			piecewise deterministic evolutions},\ }\href
	{https://doi.org/10.1103/PhysRevLett.92.120601} {\bibfield  {journal}
		{\bibinfo  {journal} {Phys. Rev. Lett.}\ }\textbf {\bibinfo {volume} {92}},\
		\bibinfo {pages} {120601} (\bibinfo {year} {2004})}\BibitemShut {NoStop}%
	\bibitem [{\citenamefont {Zhang}\ \emph {et~al.}(2022)\citenamefont {Zhang},
		\citenamefont {Yang}, \citenamefont {Guo}, \citenamefont {Sun}, \citenamefont
		{Duan}, \citenamefont {Zhou}, \citenamefont {Xie}, \citenamefont {Xu},
		\citenamefont {Gong},\ and\ \citenamefont {Zhu}}]{PhysRevA.105.042216}%
	\BibitemOpen
	\bibfield  {author} {\bibinfo {author} {\bibfnamefont {R.}~\bibnamefont
			{Zhang}}, \bibinfo {author} {\bibfnamefont {R.}~\bibnamefont {Yang}},
		\bibinfo {author} {\bibfnamefont {J.}~\bibnamefont {Guo}}, \bibinfo {author}
		{\bibfnamefont {C.-W.}\ \bibnamefont {Sun}}, \bibinfo {author} {\bibfnamefont
			{J.-C.}\ \bibnamefont {Duan}}, \bibinfo {author} {\bibfnamefont
			{H.}~\bibnamefont {Zhou}}, \bibinfo {author} {\bibfnamefont {Z.}~\bibnamefont
			{Xie}}, \bibinfo {author} {\bibfnamefont {P.}~\bibnamefont {Xu}}, \bibinfo
		{author} {\bibfnamefont {Y.-X.}\ \bibnamefont {Gong}},\ and\ \bibinfo
		{author} {\bibfnamefont {S.-N.}\ \bibnamefont {Zhu}},\ }\bibfield  {title}
	{\bibinfo {title} {Maximal coin-walker entanglement in a ballistic quantum
			walk},\ }\href {https://doi.org/10.1103/PhysRevA.105.042216} {\bibfield
		{journal} {\bibinfo  {journal} {Phys. Rev. A}\ }\textbf {\bibinfo {volume}
			{105}},\ \bibinfo {pages} {042216} (\bibinfo {year} {2022})}\BibitemShut
	{NoStop}%
	\bibitem [{\citenamefont {Widdows}\ and\ \citenamefont
		{Bhattacharyya}(2024)}]{doi:10.1177/29767032231217444}%
	\BibitemOpen
	\bibfield  {author} {\bibinfo {author} {\bibfnamefont {D.}~\bibnamefont
			{Widdows}}\ and\ \bibinfo {author} {\bibfnamefont {A.}~\bibnamefont
			{Bhattacharyya}},\ }\bibfield  {title} {\bibinfo {title} {Quantum financial
			modeling on noisy intermediate-scale quantum hardware: Random walks using
			approximate quantum counting},\ }\href
	{https://doi.org/10.1177/29767032231217444} {\bibfield  {journal} {\bibinfo
			{journal} {Quantum Economics and Finance}\ }\textbf {\bibinfo {volume} {1}},\
		\bibinfo {pages} {5} (\bibinfo {year} {2024})},\ \Eprint
	{https://arxiv.org/abs/https://doi.org/10.1177/29767032231217444}
	{https://doi.org/10.1177/29767032231217444} \BibitemShut {NoStop}%
	\bibitem [{\citenamefont {Chang}\ \emph {et~al.}(2024)\citenamefont {Chang},
		\citenamefont {Wang}, \citenamefont {Chen}, \citenamefont {Liao},\ and\
		\citenamefont {Chang}}]{Chang2024}%
	\BibitemOpen
	\bibfield  {author} {\bibinfo {author} {\bibfnamefont {Y.-J.}\ \bibnamefont
			{Chang}}, \bibinfo {author} {\bibfnamefont {W.-T.}\ \bibnamefont {Wang}},
		\bibinfo {author} {\bibfnamefont {H.-Y.}\ \bibnamefont {Chen}}, \bibinfo
		{author} {\bibfnamefont {S.-W.}\ \bibnamefont {Liao}},\ and\ \bibinfo
		{author} {\bibfnamefont {C.-R.}\ \bibnamefont {Chang}},\ }\bibfield  {title}
	{\bibinfo {title} {A novel approach for quantum financial simulation and
			quantum state preparation},\ }\href
	{https://doi.org/10.1007/s42484-024-00160-5} {\bibfield  {journal} {\bibinfo
			{journal} {Quantum Machine Intelligence}\ }\textbf {\bibinfo {volume} {6}},\
		\bibinfo {pages} {24} (\bibinfo {year} {2024})}\BibitemShut {NoStop}%
	\bibitem [{\citenamefont {Sakurai}\ and\ \citenamefont
		{Napolitano}(2010)}]{MQM}%
	\BibitemOpen
	\bibfield  {author} {\bibinfo {author} {\bibfnamefont {J.~J.}\ \bibnamefont
			{Sakurai}}\ and\ \bibinfo {author} {\bibfnamefont {J.}~\bibnamefont
			{Napolitano}},\ }\href@noop {} {\emph {\bibinfo {title} {Modern Quantum
				Mechanics}}},\ \bibinfo {edition} {2nd}\ ed.\ (\bibinfo  {publisher} {Addison
		Wesley Longman},\ \bibinfo {address} {Boston, MA},\ \bibinfo {year}
	{2010})\BibitemShut {NoStop}%
	\bibitem [{\citenamefont {Nielsen}\ and\ \citenamefont
		{Chuang}(2010)}]{Nielsen_Chuang_2010}%
	\BibitemOpen
	\bibfield  {author} {\bibinfo {author} {\bibfnamefont {M.~A.}\ \bibnamefont
			{Nielsen}}\ and\ \bibinfo {author} {\bibfnamefont {I.~L.}\ \bibnamefont
			{Chuang}},\ }\href@noop {} {\emph {\bibinfo {title} {Quantum Computation and
				Quantum Information: 10th Anniversary Edition}}}\ (\bibinfo  {publisher}
	{Cambridge University Press},\ \bibinfo {address} {Cambridge, UK},\ \bibinfo
	{year} {2010})\BibitemShut {NoStop}%
	\bibitem [{\citenamefont {Azzini}\ \emph {et~al.}(2020)\citenamefont {Azzini},
		\citenamefont {Mazzucchi}, \citenamefont {Moretti}, \citenamefont
		{Pastorello},\ and\ \citenamefont
		{Pavesi}}]{single-part-ent-10.1002-qute.202000014}%
	\BibitemOpen
	\bibfield  {author} {\bibinfo {author} {\bibfnamefont {S.}~\bibnamefont
			{Azzini}}, \bibinfo {author} {\bibfnamefont {S.}~\bibnamefont {Mazzucchi}},
		\bibinfo {author} {\bibfnamefont {V.}~\bibnamefont {Moretti}}, \bibinfo
		{author} {\bibfnamefont {D.}~\bibnamefont {Pastorello}},\ and\ \bibinfo
		{author} {\bibfnamefont {L.}~\bibnamefont {Pavesi}},\ }\bibfield  {title}
	{\bibinfo {title} {Single-particle entanglement},\ }\href
	{https://doi.org/https://doi.org/10.1002/qute.202000014} {\bibfield
		{journal} {\bibinfo  {journal} {Adv. Quantum Technol.}\ }\textbf {\bibinfo
			{volume} {3}},\ \bibinfo {pages} {2000014} (\bibinfo {year}
		{2020})}\BibitemShut {NoStop}%
	\bibitem [{\citenamefont {van Enk}(2005)}]{PhysRevA.72.064306}%
	\BibitemOpen
	\bibfield  {author} {\bibinfo {author} {\bibfnamefont {S.~J.}\ \bibnamefont
			{van Enk}},\ }\bibfield  {title} {\bibinfo {title} {Single-particle
			entanglement},\ }\href {https://doi.org/10.1103/PhysRevA.72.064306}
	{\bibfield  {journal} {\bibinfo  {journal} {Phys. Rev. A}\ }\textbf {\bibinfo
			{volume} {72}},\ \bibinfo {pages} {064306} (\bibinfo {year}
		{2005})}\BibitemShut {NoStop}%
	\bibitem [{\citenamefont {Bj\"ork}\ \emph {et~al.}(2001)\citenamefont
		{Bj\"ork}, \citenamefont {Jonsson},\ and\ \citenamefont
		{S\'anchez-Soto}}]{PhysRevA.64.042106}%
	\BibitemOpen
	\bibfield  {author} {\bibinfo {author} {\bibfnamefont {G.}~\bibnamefont
			{Bj\"ork}}, \bibinfo {author} {\bibfnamefont {P.}~\bibnamefont {Jonsson}},\
		and\ \bibinfo {author} {\bibfnamefont {L.~L.}\ \bibnamefont
			{S\'anchez-Soto}},\ }\bibfield  {title} {\bibinfo {title} {Single-particle
			nonlocality and entanglement with the vacuum},\ }\href
	{https://doi.org/10.1103/PhysRevA.64.042106} {\bibfield  {journal} {\bibinfo
			{journal} {Phys. Rev. A}\ }\textbf {\bibinfo {volume} {64}},\ \bibinfo
		{pages} {042106} (\bibinfo {year} {2001})}\BibitemShut {NoStop}%
	\bibitem [{\citenamefont {Leone}\ \emph {et~al.}(2022)\citenamefont {Leone},
		\citenamefont {Azzini}, \citenamefont {Mazzucchi}, \citenamefont {Moretti},\
		and\ \citenamefont {Pavesi}}]{PhysRevApplied.17.034011}%
	\BibitemOpen
	\bibfield  {author} {\bibinfo {author} {\bibfnamefont {N.}~\bibnamefont
			{Leone}}, \bibinfo {author} {\bibfnamefont {S.}~\bibnamefont {Azzini}},
		\bibinfo {author} {\bibfnamefont {S.}~\bibnamefont {Mazzucchi}}, \bibinfo
		{author} {\bibfnamefont {V.}~\bibnamefont {Moretti}},\ and\ \bibinfo {author}
		{\bibfnamefont {L.}~\bibnamefont {Pavesi}},\ }\bibfield  {title} {\bibinfo
		{title} {Certified quantum random-number generator based on single-photon
			entanglement},\ }\href {https://doi.org/10.1103/PhysRevApplied.17.034011}
	{\bibfield  {journal} {\bibinfo  {journal} {Phys. Rev. Appl.}\ }\textbf
		{\bibinfo {volume} {17}},\ \bibinfo {pages} {034011} (\bibinfo {year}
		{2022})}\BibitemShut {NoStop}%
	\bibitem [{\citenamefont {Lee}\ and\ \citenamefont
		{Kim}(2000)}]{PhysRevA.63.012305}%
	\BibitemOpen
	\bibfield  {author} {\bibinfo {author} {\bibfnamefont {H.-W.}\ \bibnamefont
			{Lee}}\ and\ \bibinfo {author} {\bibfnamefont {J.}~\bibnamefont {Kim}},\
	}\bibfield  {title} {\bibinfo {title} {Quantum teleportation and bell's
			inequality using single-particle entanglement},\ }\href
	{https://doi.org/10.1103/PhysRevA.63.012305} {\bibfield  {journal} {\bibinfo
			{journal} {Phys. Rev. A}\ }\textbf {\bibinfo {volume} {63}},\ \bibinfo
		{pages} {012305} (\bibinfo {year} {2000})}\BibitemShut {NoStop}%
	\bibitem [{\citenamefont {Bj\"ork}\ \emph {et~al.}(2012)\citenamefont
		{Bj\"ork}, \citenamefont {Laghaout},\ and\ \citenamefont
		{Andersen}}]{PhysRevA.85.022316}%
	\BibitemOpen
	\bibfield  {author} {\bibinfo {author} {\bibfnamefont {G.}~\bibnamefont
			{Bj\"ork}}, \bibinfo {author} {\bibfnamefont {A.}~\bibnamefont {Laghaout}},\
		and\ \bibinfo {author} {\bibfnamefont {U.~L.}\ \bibnamefont {Andersen}},\
	}\bibfield  {title} {\bibinfo {title} {Deterministic teleportation using
			single-photon entanglement as a resource},\ }\href
	{https://doi.org/10.1103/PhysRevA.85.022316} {\bibfield  {journal} {\bibinfo
			{journal} {Phys. Rev. A}\ }\textbf {\bibinfo {volume} {85}},\ \bibinfo
		{pages} {022316} (\bibinfo {year} {2012})}\BibitemShut {NoStop}%
	\bibitem [{\citenamefont {Vieira}\ \emph {et~al.}(2013)\citenamefont {Vieira},
		\citenamefont {Amorim},\ and\ \citenamefont
		{Rigolin}}]{PhysRevLett.111.180503}%
	\BibitemOpen
	\bibfield  {author} {\bibinfo {author} {\bibfnamefont {R.}~\bibnamefont
			{Vieira}}, \bibinfo {author} {\bibfnamefont {E.~P.~M.}\ \bibnamefont
			{Amorim}},\ and\ \bibinfo {author} {\bibfnamefont {G.}~\bibnamefont
			{Rigolin}},\ }\bibfield  {title} {\bibinfo {title} {Dynamically disordered
			quantum walk as a maximal entanglement generator},\ }\href
	{https://doi.org/10.1103/PhysRevLett.111.180503} {\bibfield  {journal}
		{\bibinfo  {journal} {Phys. Rev. Lett.}\ }\textbf {\bibinfo {volume} {111}},\
		\bibinfo {pages} {180503} (\bibinfo {year} {2013})}\BibitemShut {NoStop}%
	\bibitem [{\citenamefont {Orthey}\ and\ \citenamefont
		{Amorim}(2019)}]{PhysRevA.99.032320}%
	\BibitemOpen
	\bibfield  {author} {\bibinfo {author} {\bibfnamefont {A.~C.}\ \bibnamefont
			{Orthey}}\ and\ \bibinfo {author} {\bibfnamefont {E.~P.~M.}\ \bibnamefont
			{Amorim}},\ }\bibfield  {title} {\bibinfo {title} {Connecting velocity and
			entanglement in quantum walks},\ }\href
	{https://doi.org/10.1103/PhysRevA.99.032320} {\bibfield  {journal} {\bibinfo
			{journal} {Phys. Rev. A}\ }\textbf {\bibinfo {volume} {99}},\ \bibinfo
		{pages} {032320} (\bibinfo {year} {2019})}\BibitemShut {NoStop}%
	\bibitem [{\citenamefont {Vakulchyk}\ \emph {et~al.}(2017)\citenamefont
		{Vakulchyk}, \citenamefont {Fistul}, \citenamefont {Qin},\ and\ \citenamefont
		{Flach}}]{PhysRevB.96.144204}%
	\BibitemOpen
	\bibfield  {author} {\bibinfo {author} {\bibfnamefont {I.}~\bibnamefont
			{Vakulchyk}}, \bibinfo {author} {\bibfnamefont {M.~V.}\ \bibnamefont
			{Fistul}}, \bibinfo {author} {\bibfnamefont {P.}~\bibnamefont {Qin}},\ and\
		\bibinfo {author} {\bibfnamefont {S.}~\bibnamefont {Flach}},\ }\bibfield
	{title} {\bibinfo {title} {Anderson localization in generalized discrete-time
			quantum walks},\ }\href {https://doi.org/10.1103/PhysRevB.96.144204}
	{\bibfield  {journal} {\bibinfo  {journal} {Phys. Rev. B}\ }\textbf {\bibinfo
			{volume} {96}},\ \bibinfo {pages} {144204} (\bibinfo {year}
		{2017})}\BibitemShut {NoStop}%
	\bibitem [{\citenamefont {Schreiber}\ \emph {et~al.}(2011)\citenamefont
		{Schreiber}, \citenamefont {Cassemiro}, \citenamefont
		{Poto\ifmmode~\check{c}\else \v{c}\fi{}ek}, \citenamefont {G\'abris},
		\citenamefont {Jex},\ and\ \citenamefont
		{Silberhorn}}]{PhysRevLett.106.180403}%
	\BibitemOpen
	\bibfield  {author} {\bibinfo {author} {\bibfnamefont {A.}~\bibnamefont
			{Schreiber}}, \bibinfo {author} {\bibfnamefont {K.~N.}\ \bibnamefont
			{Cassemiro}}, \bibinfo {author} {\bibfnamefont {V.}~\bibnamefont
			{Poto\ifmmode~\check{c}\else \v{c}\fi{}ek}}, \bibinfo {author} {\bibfnamefont
			{A.}~\bibnamefont {G\'abris}}, \bibinfo {author} {\bibfnamefont
			{I.}~\bibnamefont {Jex}},\ and\ \bibinfo {author} {\bibfnamefont
			{C.}~\bibnamefont {Silberhorn}},\ }\bibfield  {title} {\bibinfo {title}
		{Decoherence and disorder in quantum walks: From ballistic spread to
			localization},\ }\href {https://doi.org/10.1103/PhysRevLett.106.180403}
	{\bibfield  {journal} {\bibinfo  {journal} {Phys. Rev. Lett.}\ }\textbf
		{\bibinfo {volume} {106}},\ \bibinfo {pages} {180403} (\bibinfo {year}
		{2011})}\BibitemShut {NoStop}%
	\bibitem [{\citenamefont {Malishava}\ \emph {et~al.}(2020)\citenamefont
		{Malishava}, \citenamefont {Vakulchyk}, \citenamefont {Fistul},\ and\
		\citenamefont {Flach}}]{PhysRevB.101.144201}%
	\BibitemOpen
	\bibfield  {author} {\bibinfo {author} {\bibfnamefont {M.}~\bibnamefont
			{Malishava}}, \bibinfo {author} {\bibfnamefont {I.}~\bibnamefont
			{Vakulchyk}}, \bibinfo {author} {\bibfnamefont {M.}~\bibnamefont {Fistul}},\
		and\ \bibinfo {author} {\bibfnamefont {S.}~\bibnamefont {Flach}},\ }\bibfield
	{title} {\bibinfo {title} {Floquet anderson localization of two interacting
			discrete time quantum walks},\ }\href
	{https://doi.org/10.1103/PhysRevB.101.144201} {\bibfield  {journal} {\bibinfo
			{journal} {Phys. Rev. B}\ }\textbf {\bibinfo {volume} {101}},\ \bibinfo
		{pages} {144201} (\bibinfo {year} {2020})}\BibitemShut {NoStop}%
	\bibitem [{\citenamefont {Yao}\ and\ \citenamefont
		{Wald}(2023)}]{PhysRevE.108.024139}%
	\BibitemOpen
	\bibfield  {author} {\bibinfo {author} {\bibfnamefont {L.~H.}\ \bibnamefont
			{Yao}}\ and\ \bibinfo {author} {\bibfnamefont {S.}~\bibnamefont {Wald}},\
	}\bibfield  {title} {\bibinfo {title} {Coined quantum walks on the line:
			Disorder, entanglement, and localization},\ }\href
	{https://doi.org/10.1103/PhysRevE.108.024139} {\bibfield  {journal} {\bibinfo
			{journal} {Phys. Rev. E}\ }\textbf {\bibinfo {volume} {108}},\ \bibinfo
		{pages} {024139} (\bibinfo {year} {2023})}\BibitemShut {NoStop}%
	\bibitem [{\citenamefont {Rakovszky}\ and\ \citenamefont
		{Asboth}(2015)}]{PhysRevA.92.052311}%
	\BibitemOpen
	\bibfield  {author} {\bibinfo {author} {\bibfnamefont {T.}~\bibnamefont
			{Rakovszky}}\ and\ \bibinfo {author} {\bibfnamefont {J.~K.}\ \bibnamefont
			{Asboth}},\ }\bibfield  {title} {\bibinfo {title} {Localization,
			delocalization, and topological phase transitions in the one-dimensional
			split-step quantum walk},\ }\href
	{https://doi.org/10.1103/PhysRevA.92.052311} {\bibfield  {journal} {\bibinfo
			{journal} {Phys. Rev. A}\ }\textbf {\bibinfo {volume} {92}},\ \bibinfo
		{pages} {052311} (\bibinfo {year} {2015})}\BibitemShut {NoStop}%
	\bibitem [{\citenamefont {Anderson}(1958)}]{PhysRev.109.1492}%
	\BibitemOpen
	\bibfield  {author} {\bibinfo {author} {\bibfnamefont {P.~W.}\ \bibnamefont
			{Anderson}},\ }\bibfield  {title} {\bibinfo {title} {Absence of diffusion in
			certain random lattices},\ }\href {https://doi.org/10.1103/PhysRev.109.1492}
	{\bibfield  {journal} {\bibinfo  {journal} {Phys. Rev.}\ }\textbf {\bibinfo
			{volume} {109}},\ \bibinfo {pages} {1492} (\bibinfo {year}
		{1958})}\BibitemShut {NoStop}%
	\bibitem [{\citenamefont {Song}\ \emph {et~al.}(2011)\citenamefont {Song},
		\citenamefont {Flindt}, \citenamefont {Rachel}, \citenamefont {Klich},\ and\
		\citenamefont {Le~Hur}}]{PhysRevB.83.161408}%
	\BibitemOpen
	\bibfield  {author} {\bibinfo {author} {\bibfnamefont {H.~F.}\ \bibnamefont
			{Song}}, \bibinfo {author} {\bibfnamefont {C.}~\bibnamefont {Flindt}},
		\bibinfo {author} {\bibfnamefont {S.}~\bibnamefont {Rachel}}, \bibinfo
		{author} {\bibfnamefont {I.}~\bibnamefont {Klich}},\ and\ \bibinfo {author}
		{\bibfnamefont {K.}~\bibnamefont {Le~Hur}},\ }\bibfield  {title} {\bibinfo
		{title} {Entanglement entropy from charge statistics: Exact relations for
			noninteracting many-body systems},\ }\href
	{https://doi.org/10.1103/PhysRevB.83.161408} {\bibfield  {journal} {\bibinfo
			{journal} {Phys. Rev. B}\ }\textbf {\bibinfo {volume} {83}},\ \bibinfo
		{pages} {161408} (\bibinfo {year} {2011})}\BibitemShut {NoStop}%
	\bibitem [{\citenamefont {Islam}\ \emph {et~al.}(2015)\citenamefont {Islam},
		\citenamefont {Ma}, \citenamefont {Preiss}, \citenamefont {Eric~Tai},
		\citenamefont {Lukin}, \citenamefont {Rispoli},\ and\ \citenamefont
		{Greiner}}]{Islam2015}%
	\BibitemOpen
	\bibfield  {author} {\bibinfo {author} {\bibfnamefont {R.}~\bibnamefont
			{Islam}}, \bibinfo {author} {\bibfnamefont {R.}~\bibnamefont {Ma}}, \bibinfo
		{author} {\bibfnamefont {P.~M.}\ \bibnamefont {Preiss}}, \bibinfo {author}
		{\bibfnamefont {M.}~\bibnamefont {Eric~Tai}}, \bibinfo {author}
		{\bibfnamefont {A.}~\bibnamefont {Lukin}}, \bibinfo {author} {\bibfnamefont
			{M.}~\bibnamefont {Rispoli}},\ and\ \bibinfo {author} {\bibfnamefont
			{M.}~\bibnamefont {Greiner}},\ }\bibfield  {title} {\bibinfo {title}
		{Measuring entanglement entropy in a quantum many-body system},\ }\href
	{https://doi.org/10.1038/nature15750} {\bibfield  {journal} {\bibinfo
			{journal} {Nature}\ }\textbf {\bibinfo {volume} {528}},\ \bibinfo {pages}
		{77} (\bibinfo {year} {2015})}\BibitemShut {NoStop}%
	\bibitem [{\citenamefont {Kaufman}\ \emph {et~al.}(2016)\citenamefont
		{Kaufman}, \citenamefont {Tai}, \citenamefont {Lukin}, \citenamefont
		{Rispoli}, \citenamefont {Schittko}, \citenamefont {Preiss},\ and\
		\citenamefont {Greiner}}]{Kaufman16}%
	\BibitemOpen
	\bibfield  {author} {\bibinfo {author} {\bibfnamefont {A.~M.}\ \bibnamefont
			{Kaufman}}, \bibinfo {author} {\bibfnamefont {M.~E.}\ \bibnamefont {Tai}},
		\bibinfo {author} {\bibfnamefont {A.}~\bibnamefont {Lukin}}, \bibinfo
		{author} {\bibfnamefont {M.}~\bibnamefont {Rispoli}}, \bibinfo {author}
		{\bibfnamefont {R.}~\bibnamefont {Schittko}}, \bibinfo {author}
		{\bibfnamefont {P.~M.}\ \bibnamefont {Preiss}},\ and\ \bibinfo {author}
		{\bibfnamefont {M.}~\bibnamefont {Greiner}},\ }\bibfield  {title} {\bibinfo
		{title} {Quantum thermalization through entanglement in an isolated many-body
			system},\ }\href {https://doi.org/10.1126/science.aaf6725} {\bibfield
		{journal} {\bibinfo  {journal} {Science}\ }\textbf {\bibinfo {volume}
			{353}},\ \bibinfo {pages} {794} (\bibinfo {year} {2016})},\ \Eprint
	{https://arxiv.org/abs/https://www.science.org/doi/pdf/10.1126/science.aaf6725}
	{https://www.science.org/doi/pdf/10.1126/science.aaf6725} \BibitemShut
	{NoStop}%
	\bibitem [{\citenamefont {Daley}\ \emph {et~al.}(2012)\citenamefont {Daley},
		\citenamefont {Pichler}, \citenamefont {Schachenmayer},\ and\ \citenamefont
		{Zoller}}]{PhysRevLett.109.020505}%
	\BibitemOpen
	\bibfield  {author} {\bibinfo {author} {\bibfnamefont {A.~J.}\ \bibnamefont
			{Daley}}, \bibinfo {author} {\bibfnamefont {H.}~\bibnamefont {Pichler}},
		\bibinfo {author} {\bibfnamefont {J.}~\bibnamefont {Schachenmayer}},\ and\
		\bibinfo {author} {\bibfnamefont {P.}~\bibnamefont {Zoller}},\ }\bibfield
	{title} {\bibinfo {title} {Measuring entanglement growth in quench dynamics
			of bosons in an optical lattice},\ }\href
	{https://doi.org/10.1103/PhysRevLett.109.020505} {\bibfield  {journal}
		{\bibinfo  {journal} {Phys. Rev. Lett.}\ }\textbf {\bibinfo {volume} {109}},\
		\bibinfo {pages} {020505} (\bibinfo {year} {2012})}\BibitemShut {NoStop}%
	\bibitem [{\citenamefont {Han}\ \emph {et~al.}(2023)\citenamefont {Han},
		\citenamefont {Meir},\ and\ \citenamefont {Sela}}]{PhysRevLett.130.136201}%
	\BibitemOpen
	\bibfield  {author} {\bibinfo {author} {\bibfnamefont {C.}~\bibnamefont
			{Han}}, \bibinfo {author} {\bibfnamefont {Y.}~\bibnamefont {Meir}},\ and\
		\bibinfo {author} {\bibfnamefont {E.}~\bibnamefont {Sela}},\ }\bibfield
	{title} {\bibinfo {title} {Realistic protocol to measure entanglement at
			finite temperatures},\ }\href
	{https://doi.org/10.1103/PhysRevLett.130.136201} {\bibfield  {journal}
		{\bibinfo  {journal} {Phys. Rev. Lett.}\ }\textbf {\bibinfo {volume} {130}},\
		\bibinfo {pages} {136201} (\bibinfo {year} {2023})}\BibitemShut {NoStop}%
	\bibitem [{\citenamefont {Lin}\ \emph {et~al.}(2024)\citenamefont {Lin},
		\citenamefont {Zhou}, \citenamefont {Jiang}, \citenamefont {Wu},
		\citenamefont {Chen}, \citenamefont {Liu}, \citenamefont {Wang},
		\citenamefont {Ye},\ and\ \citenamefont {Jiang}}]{Lin2024}%
	\BibitemOpen
	\bibfield  {author} {\bibinfo {author} {\bibfnamefont {Z.-K.}\ \bibnamefont
			{Lin}}, \bibinfo {author} {\bibfnamefont {Y.}~\bibnamefont {Zhou}}, \bibinfo
		{author} {\bibfnamefont {B.}~\bibnamefont {Jiang}}, \bibinfo {author}
		{\bibfnamefont {B.-Q.}\ \bibnamefont {Wu}}, \bibinfo {author} {\bibfnamefont
			{L.-M.}\ \bibnamefont {Chen}}, \bibinfo {author} {\bibfnamefont {X.-Y.}\
			\bibnamefont {Liu}}, \bibinfo {author} {\bibfnamefont {L.-W.}\ \bibnamefont
			{Wang}}, \bibinfo {author} {\bibfnamefont {P.}~\bibnamefont {Ye}},\ and\
		\bibinfo {author} {\bibfnamefont {J.-H.}\ \bibnamefont {Jiang}},\ }\bibfield
	{title} {\bibinfo {title} {Measuring entanglement entropy and its topological
			signature for phononic systems},\ }\href
	{https://doi.org/10.1038/s41467-024-45887-8} {\bibfield  {journal} {\bibinfo
			{journal} {Nature Communications}\ }\textbf {\bibinfo {volume} {15}},\
		\bibinfo {pages} {1601} (\bibinfo {year} {2024})}\BibitemShut {NoStop}%
	\bibitem [{\citenamefont {Khoo}\ and\ \citenamefont
		{Heyl}(2021)}]{khoo_quantum_2021}%
	\BibitemOpen
	\bibfield  {author} {\bibinfo {author} {\bibfnamefont {J.~Y.}\ \bibnamefont
			{Khoo}}\ and\ \bibinfo {author} {\bibfnamefont {M.}~\bibnamefont {Heyl}},\
	}\bibfield  {title} {\bibinfo {title} {Quantum entanglement recognition},\
	}\href {https://doi.org/10.1103/PhysRevResearch.3.033135} {\bibfield
		{journal} {\bibinfo  {journal} {Phys. Rev. Res.}\ }\textbf {\bibinfo {volume}
			{3}},\ \bibinfo {pages} {033135} (\bibinfo {year} {2021})}\BibitemShut
	{NoStop}%
	\bibitem [{\citenamefont {Wang}\ \emph {et~al.}(2022)\citenamefont {Wang},
		\citenamefont {Tang}, \citenamefont {Chen},\ and\ \citenamefont
		{Zhang}}]{doi.org-10.100-lpor.202100519}%
	\BibitemOpen
	\bibfield  {author} {\bibinfo {author} {\bibfnamefont {B.}~\bibnamefont
			{Wang}}, \bibinfo {author} {\bibfnamefont {Z.}~\bibnamefont {Tang}}, \bibinfo
		{author} {\bibfnamefont {T.}~\bibnamefont {Chen}},\ and\ \bibinfo {author}
		{\bibfnamefont {X.}~\bibnamefont {Zhang}},\ }\bibfield  {title} {\bibinfo
		{title} {Nearly perfect transmission and transformation of entangled states
			in topologically protected channels},\ }\href
	{https://doi.org/https://doi.org/10.1002/lpor.202100519} {\bibfield
		{journal} {\bibinfo  {journal} {Laser \& Photonics Reviews}\ }\textbf
		{\bibinfo {volume} {16}},\ \bibinfo {pages} {2100519} (\bibinfo {year}
		{2022})},\ \Eprint
	{https://arxiv.org/abs/https://onlinelibrary.wiley.com/doi/pdf/10.1002/lpor.202100519}
	{https://onlinelibrary.wiley.com/doi/pdf/10.1002/lpor.202100519} \BibitemShut
	{NoStop}%
	\bibitem [{\citenamefont {Tang}\ \emph {et~al.}(2024)\citenamefont {Tang},
		\citenamefont {Chen}, \citenamefont {Tang},\ and\ \citenamefont
		{Zhang}}]{Tang2024}%
	\BibitemOpen
	\bibfield  {author} {\bibinfo {author} {\bibfnamefont {Z.}~\bibnamefont
			{Tang}}, \bibinfo {author} {\bibfnamefont {T.}~\bibnamefont {Chen}}, \bibinfo
		{author} {\bibfnamefont {X.}~\bibnamefont {Tang}},\ and\ \bibinfo {author}
		{\bibfnamefont {X.}~\bibnamefont {Zhang}},\ }\bibfield  {title} {\bibinfo
		{title} {Topologically protected entanglement switching around exceptional
			points},\ }\href {https://doi.org/10.1038/s41377-024-01514-1} {\bibfield
		{journal} {\bibinfo  {journal} {Light: Science {\&} Applications}\ }\textbf
		{\bibinfo {volume} {13}},\ \bibinfo {pages} {167} (\bibinfo {year}
		{2024})}\BibitemShut {NoStop}%
	\bibitem [{\citenamefont {Tang}\ \emph {et~al.}(2022)\citenamefont {Tang},
		\citenamefont {Wang}, \citenamefont {Chen},\ and\ \citenamefont
		{Zhang}}]{PhysRevResearch.4.043144}%
	\BibitemOpen
	\bibfield  {author} {\bibinfo {author} {\bibfnamefont {Z.}~\bibnamefont
			{Tang}}, \bibinfo {author} {\bibfnamefont {B.}~\bibnamefont {Wang}}, \bibinfo
		{author} {\bibfnamefont {T.}~\bibnamefont {Chen}},\ and\ \bibinfo {author}
		{\bibfnamefont {X.}~\bibnamefont {Zhang}},\ }\bibfield  {title} {\bibinfo
		{title} {Transmission and transformation of entangled states with high
			fidelity in a non-hermitian system},\ }\href
	{https://doi.org/10.1103/PhysRevResearch.4.043144} {\bibfield  {journal}
		{\bibinfo  {journal} {Phys. Rev. Res.}\ }\textbf {\bibinfo {volume} {4}},\
		\bibinfo {pages} {043144} (\bibinfo {year} {2022})}\BibitemShut {NoStop}%
	\bibitem [{\citenamefont {Omar}\ \emph {et~al.}(2006)\citenamefont {Omar},
		\citenamefont {Paunkovi\ifmmode~\acute{c}\else \'{c}\fi{}}, \citenamefont
		{Sheridan},\ and\ \citenamefont {Bose}}]{PhysRevA.74.042304}%
	\BibitemOpen
	\bibfield  {author} {\bibinfo {author} {\bibfnamefont {Y.}~\bibnamefont
			{Omar}}, \bibinfo {author} {\bibfnamefont {N.}~\bibnamefont
			{Paunkovi\ifmmode~\acute{c}\else \'{c}\fi{}}}, \bibinfo {author}
		{\bibfnamefont {L.}~\bibnamefont {Sheridan}},\ and\ \bibinfo {author}
		{\bibfnamefont {S.}~\bibnamefont {Bose}},\ }\bibfield  {title} {\bibinfo
		{title} {Quantum walk on a line with two entangled particles},\ }\href
	{https://doi.org/10.1103/PhysRevA.74.042304} {\bibfield  {journal} {\bibinfo
			{journal} {Phys. Rev. A}\ }\textbf {\bibinfo {volume} {74}},\ \bibinfo
		{pages} {042304} (\bibinfo {year} {2006})}\BibitemShut {NoStop}%
	\bibitem [{\citenamefont {Francisco}\ \emph {et~al.}(2006)\citenamefont
		{Francisco}, \citenamefont {Iemmi}, \citenamefont {Paz},\ and\ \citenamefont
		{Ledesma}}]{PhysRevA.74.052327}%
	\BibitemOpen
	\bibfield  {author} {\bibinfo {author} {\bibfnamefont {D.}~\bibnamefont
			{Francisco}}, \bibinfo {author} {\bibfnamefont {C.}~\bibnamefont {Iemmi}},
		\bibinfo {author} {\bibfnamefont {J.~P.}\ \bibnamefont {Paz}},\ and\ \bibinfo
		{author} {\bibfnamefont {S.}~\bibnamefont {Ledesma}},\ }\bibfield  {title}
	{\bibinfo {title} {Simulating a quantum walk with classical optics},\ }\href
	{https://doi.org/10.1103/PhysRevA.74.052327} {\bibfield  {journal} {\bibinfo
			{journal} {Phys. Rev. A}\ }\textbf {\bibinfo {volume} {74}},\ \bibinfo
		{pages} {052327} (\bibinfo {year} {2006})}\BibitemShut {NoStop}%
	\bibitem [{\citenamefont {Mastandrea}\ and\ \citenamefont
		{Chien}(2023)}]{PhysRevE.108.035308}%
	\BibitemOpen
	\bibfield  {author} {\bibinfo {author} {\bibfnamefont {C.}~\bibnamefont
			{Mastandrea}}\ and\ \bibinfo {author} {\bibfnamefont {C.-C.}\ \bibnamefont
			{Chien}},\ }\bibfield  {title} {\bibinfo {title} {Localization of quantum
			walks with classical randomness: Comparison between manual methods and
			supervised machine learning},\ }\href
	{https://doi.org/10.1103/PhysRevE.108.035308} {\bibfield  {journal} {\bibinfo
			{journal} {Phys. Rev. E}\ }\textbf {\bibinfo {volume} {108}},\ \bibinfo
		{pages} {035308} (\bibinfo {year} {2023})}\BibitemShut {NoStop}%
	\bibitem [{\citenamefont {Chandrashekar}\ and\ \citenamefont
		{Laflamme}(2008)}]{PhysRevA.78.022314}%
	\BibitemOpen
	\bibfield  {author} {\bibinfo {author} {\bibfnamefont {C.~M.}\ \bibnamefont
			{Chandrashekar}}\ and\ \bibinfo {author} {\bibfnamefont {R.}~\bibnamefont
			{Laflamme}},\ }\bibfield  {title} {\bibinfo {title} {Quantum phase transition
			using quantum walks in an optical lattice},\ }\href
	{https://doi.org/10.1103/PhysRevA.78.022314} {\bibfield  {journal} {\bibinfo
			{journal} {Phys. Rev. A}\ }\textbf {\bibinfo {volume} {78}},\ \bibinfo
		{pages} {022314} (\bibinfo {year} {2008})}\BibitemShut {NoStop}%
	\bibitem [{\citenamefont {Kitagawa}\ \emph {et~al.}(2010)\citenamefont
		{Kitagawa}, \citenamefont {Rudner}, \citenamefont {Berg},\ and\ \citenamefont
		{Demler}}]{PhysRevA.82.033429}%
	\BibitemOpen
	\bibfield  {author} {\bibinfo {author} {\bibfnamefont {T.}~\bibnamefont
			{Kitagawa}}, \bibinfo {author} {\bibfnamefont {M.~S.}\ \bibnamefont
			{Rudner}}, \bibinfo {author} {\bibfnamefont {E.}~\bibnamefont {Berg}},\ and\
		\bibinfo {author} {\bibfnamefont {E.}~\bibnamefont {Demler}},\ }\bibfield
	{title} {\bibinfo {title} {Exploring topological phases with quantum walks},\
	}\href {https://doi.org/10.1103/PhysRevA.82.033429} {\bibfield  {journal}
		{\bibinfo  {journal} {Phys. Rev. A}\ }\textbf {\bibinfo {volume} {82}},\
		\bibinfo {pages} {033429} (\bibinfo {year} {2010})}\BibitemShut {NoStop}%
	\bibitem [{\citenamefont {Tarasinski}\ \emph {et~al.}(2014)\citenamefont
		{Tarasinski}, \citenamefont {Asb\'oth},\ and\ \citenamefont
		{Dahlhaus}}]{PhysRevA.89.042327}%
	\BibitemOpen
	\bibfield  {author} {\bibinfo {author} {\bibfnamefont {B.}~\bibnamefont
			{Tarasinski}}, \bibinfo {author} {\bibfnamefont {J.~K.}\ \bibnamefont
			{Asb\'oth}},\ and\ \bibinfo {author} {\bibfnamefont {J.~P.}\ \bibnamefont
			{Dahlhaus}},\ }\bibfield  {title} {\bibinfo {title} {Scattering theory of
			topological phases in discrete-time quantum walks},\ }\href
	{https://doi.org/10.1103/PhysRevA.89.042327} {\bibfield  {journal} {\bibinfo
			{journal} {Phys. Rev. A}\ }\textbf {\bibinfo {volume} {89}},\ \bibinfo
		{pages} {042327} (\bibinfo {year} {2014})}\BibitemShut {NoStop}%
	\bibitem [{\citenamefont {Peng}\ \emph {et~al.}(2021)\citenamefont {Peng},
		\citenamefont {Wang},\ and\ \citenamefont {Yi}}]{PhysRevA.103.032205}%
	\BibitemOpen
	\bibfield  {author} {\bibinfo {author} {\bibfnamefont {Y.~F.}\ \bibnamefont
			{Peng}}, \bibinfo {author} {\bibfnamefont {W.}~\bibnamefont {Wang}},\ and\
		\bibinfo {author} {\bibfnamefont {X.~X.}\ \bibnamefont {Yi}},\ }\bibfield
	{title} {\bibinfo {title} {Discrete-time quantum walk with time-correlated
			noise},\ }\href {https://doi.org/10.1103/PhysRevA.103.032205} {\bibfield
		{journal} {\bibinfo  {journal} {Phys. Rev. A}\ }\textbf {\bibinfo {volume}
			{103}},\ \bibinfo {pages} {032205} (\bibinfo {year} {2021})}\BibitemShut
	{NoStop}%
	\bibitem [{\citenamefont {Wojcik}\ \emph {et~al.}(2012)\citenamefont {Wojcik},
		\citenamefont {Luczak}, \citenamefont {Kurzynski}, \citenamefont {Grudka},
		\citenamefont {Gdala},\ and\ \citenamefont
		{Bednarska-Bzdega}}]{PhysRevA.85.012329}%
	\BibitemOpen
	\bibfield  {author} {\bibinfo {author} {\bibfnamefont {A.}~\bibnamefont
			{Wojcik}}, \bibinfo {author} {\bibfnamefont {T.}~\bibnamefont {Luczak}},
		\bibinfo {author} {\bibfnamefont {P.}~\bibnamefont {Kurzynski}}, \bibinfo
		{author} {\bibfnamefont {A.}~\bibnamefont {Grudka}}, \bibinfo {author}
		{\bibfnamefont {T.}~\bibnamefont {Gdala}},\ and\ \bibinfo {author}
		{\bibfnamefont {M.}~\bibnamefont {Bednarska-Bzdega}},\ }\bibfield  {title}
	{\bibinfo {title} {Trapping a particle of a quantum walk on the line},\
	}\href {https://doi.org/10.1103/PhysRevA.85.012329} {\bibfield  {journal}
		{\bibinfo  {journal} {Phys. Rev. A}\ }\textbf {\bibinfo {volume} {85}},\
		\bibinfo {pages} {012329} (\bibinfo {year} {2012})}\BibitemShut {NoStop}%
	\bibitem [{\citenamefont {Ishak}\ \emph {et~al.}(2021)\citenamefont {Ishak},
		\citenamefont {Muniandy},\ and\ \citenamefont {Chong}}]{ISHAK2021126371}%
	\BibitemOpen
	\bibfield  {author} {\bibinfo {author} {\bibfnamefont {N.~I.}\ \bibnamefont
			{Ishak}}, \bibinfo {author} {\bibfnamefont {S.}~\bibnamefont {Muniandy}},\
		and\ \bibinfo {author} {\bibfnamefont {W.~Y.}\ \bibnamefont {Chong}},\
	}\bibfield  {title} {\bibinfo {title} {Entropy analysis of the discrete-time
			quantum walk under bit-flip noise channel},\ }\href
	{https://doi.org/https://doi.org/10.1016/j.physa.2021.126371} {\bibfield
		{journal} {\bibinfo  {journal} {Physica A: Statistical Mechanics and its
				Applications}\ }\textbf {\bibinfo {volume} {584}},\ \bibinfo {pages} {126371}
		(\bibinfo {year} {2021})}\BibitemShut {NoStop}%
	\bibitem [{\citenamefont {Mittal}\ and\ \citenamefont
		{Huang}(2024)}]{Mittal24}%
	\BibitemOpen
	\bibfield  {author} {\bibinfo {author} {\bibfnamefont {V.}~\bibnamefont
			{Mittal}}\ and\ \bibinfo {author} {\bibfnamefont {Y.~P.}\ \bibnamefont
			{Huang}},\ }\href@noop {} {\bibinfo {title} {Parrondo’s paradox in quantum
			walks with inhomogeneous coins}} (\bibinfo {year} {2024}),\ \bibinfo {note}
	{arXiv: 2407.16558}\BibitemShut {NoStop}%
	\bibitem [{\citenamefont {Danaci}\ \emph {et~al.}(2021)\citenamefont {Danaci},
		\citenamefont {Yalcinkaya}, \citenamefont {Cakmak}, \citenamefont {Karpat},
		\citenamefont {Kelly},\ and\ \citenamefont {Subasi}}]{PhysRevA.103.022416}%
	\BibitemOpen
	\bibfield  {author} {\bibinfo {author} {\bibfnamefont {B.}~\bibnamefont
			{Danaci}}, \bibinfo {author} {\bibfnamefont {I.}~\bibnamefont {Yalcinkaya}},
		\bibinfo {author} {\bibfnamefont {B.}~\bibnamefont {Cakmak}}, \bibinfo
		{author} {\bibfnamefont {G.}~\bibnamefont {Karpat}}, \bibinfo {author}
		{\bibfnamefont {S.~P.}\ \bibnamefont {Kelly}},\ and\ \bibinfo {author}
		{\bibfnamefont {A.~L.}\ \bibnamefont {Subasi}},\ }\bibfield  {title}
	{\bibinfo {title} {Disorder-free localization in quantum walks},\ }\href
	{https://doi.org/10.1103/PhysRevA.103.022416} {\bibfield  {journal} {\bibinfo
			{journal} {Phys. Rev. A}\ }\textbf {\bibinfo {volume} {103}},\ \bibinfo
		{pages} {022416} (\bibinfo {year} {2021})}\BibitemShut {NoStop}%
	\bibitem [{\citenamefont {Chien}\ \emph {et~al.}(2014)\citenamefont {Chien},
		\citenamefont {Di~Ventra},\ and\ \citenamefont
		{Zwolak}}]{PhysRevA.90.023624}%
	\BibitemOpen
	\bibfield  {author} {\bibinfo {author} {\bibfnamefont {C.-C.}\ \bibnamefont
			{Chien}}, \bibinfo {author} {\bibfnamefont {M.}~\bibnamefont {Di~Ventra}},\
		and\ \bibinfo {author} {\bibfnamefont {M.}~\bibnamefont {Zwolak}},\
	}\bibfield  {title} {\bibinfo {title} {Landauer, kubo, and microcanonical
			approaches to quantum transport and noise: A comparison and implications for
			cold-atom dynamics},\ }\href {https://doi.org/10.1103/PhysRevA.90.023624}
	{\bibfield  {journal} {\bibinfo  {journal} {Phys. Rev. A}\ }\textbf {\bibinfo
			{volume} {90}},\ \bibinfo {pages} {023624} (\bibinfo {year}
		{2014})}\BibitemShut {NoStop}%
	\bibitem [{\citenamefont {Travaglione}\ and\ \citenamefont
		{Milburn}(2002)}]{PhysRevA.65.032310}%
	\BibitemOpen
	\bibfield  {author} {\bibinfo {author} {\bibfnamefont {B.~C.}\ \bibnamefont
			{Travaglione}}\ and\ \bibinfo {author} {\bibfnamefont {G.~J.}\ \bibnamefont
			{Milburn}},\ }\bibfield  {title} {\bibinfo {title} {Implementing the quantum
			random walk},\ }\href {https://doi.org/10.1103/PhysRevA.65.032310} {\bibfield
		{journal} {\bibinfo  {journal} {Phys. Rev. A}\ }\textbf {\bibinfo {volume}
			{65}},\ \bibinfo {pages} {032310} (\bibinfo {year} {2002})}\BibitemShut
	{NoStop}%
	\bibitem [{\citenamefont {Jayakody}\ \emph {et~al.}(2023)\citenamefont
		{Jayakody}, \citenamefont {Meena},\ and\ \citenamefont
		{Pradhan}}]{JAYAKODY2023100189}%
	\BibitemOpen
	\bibfield  {author} {\bibinfo {author} {\bibfnamefont {M.~N.}\ \bibnamefont
			{Jayakody}}, \bibinfo {author} {\bibfnamefont {C.}~\bibnamefont {Meena}},\
		and\ \bibinfo {author} {\bibfnamefont {P.}~\bibnamefont {Pradhan}},\
	}\bibfield  {title} {\bibinfo {title} {Revisiting one-dimensional
			discrete-time quantum walks with general coin},\ }\href
	{https://doi.org/https://doi.org/10.1016/j.physo.2023.100189} {\bibfield
		{journal} {\bibinfo  {journal} {Physics Open}\ }\textbf {\bibinfo {volume}
			{17}},\ \bibinfo {pages} {100189} (\bibinfo {year} {2023})}\BibitemShut
	{NoStop}%
	\bibitem [{\citenamefont {Gratsea}\ \emph {et~al.}(2020)\citenamefont
		{Gratsea}, \citenamefont {Metz},\ and\ \citenamefont {Busch}}]{Gratsea_2020}%
	\BibitemOpen
	\bibfield  {author} {\bibinfo {author} {\bibfnamefont {A.}~\bibnamefont
			{Gratsea}}, \bibinfo {author} {\bibfnamefont {F.}~\bibnamefont {Metz}},\ and\
		\bibinfo {author} {\bibfnamefont {T.}~\bibnamefont {Busch}},\ }\bibfield
	{title} {\bibinfo {title} {Universal and optimal coin sequences for high
			entanglement generation in 1d discrete time quantum walks},\ }\href
	{https://doi.org/10.1088/1751-8121/abb54d} {\bibfield  {journal} {\bibinfo
			{journal} {J. Phys. A: Math. Theor.}\ }\textbf {\bibinfo {volume} {53}},\
		\bibinfo {pages} {445306} (\bibinfo {year} {2020})}\BibitemShut {NoStop}%
	\bibitem [{\citenamefont {Brun}\ \emph {et~al.}(2003)\citenamefont {Brun},
		\citenamefont {Carteret},\ and\ \citenamefont
		{Ambainis}}]{PhysRevA.67.052317}%
	\BibitemOpen
	\bibfield  {author} {\bibinfo {author} {\bibfnamefont {T.~A.}\ \bibnamefont
			{Brun}}, \bibinfo {author} {\bibfnamefont {H.~A.}\ \bibnamefont {Carteret}},\
		and\ \bibinfo {author} {\bibfnamefont {A.}~\bibnamefont {Ambainis}},\
	}\bibfield  {title} {\bibinfo {title} {Quantum walks driven by many coins},\
	}\href {https://doi.org/10.1103/PhysRevA.67.052317} {\bibfield  {journal}
		{\bibinfo  {journal} {Phys. Rev. A}\ }\textbf {\bibinfo {volume} {67}},\
		\bibinfo {pages} {052317} (\bibinfo {year} {2003})}\BibitemShut {NoStop}%
	\bibitem [{\citenamefont {Su}\ \emph {et~al.}(2019)\citenamefont {Su},
		\citenamefont {Zhang}, \citenamefont {Yu}, \citenamefont {Zhou},
		\citenamefont {Jin}, \citenamefont {Xu}, \citenamefont {Xiong}, \citenamefont
		{Xu}, \citenamefont {Sun}, \citenamefont {Chen}, \citenamefont {Nori},\ and\
		\citenamefont {Yang}}]{Su2019}%
	\BibitemOpen
	\bibfield  {author} {\bibinfo {author} {\bibfnamefont {Q.-P.}\ \bibnamefont
			{Su}}, \bibinfo {author} {\bibfnamefont {Y.}~\bibnamefont {Zhang}}, \bibinfo
		{author} {\bibfnamefont {L.}~\bibnamefont {Yu}}, \bibinfo {author}
		{\bibfnamefont {J.-Q.}\ \bibnamefont {Zhou}}, \bibinfo {author}
		{\bibfnamefont {J.-S.}\ \bibnamefont {Jin}}, \bibinfo {author} {\bibfnamefont
			{X.-Q.}\ \bibnamefont {Xu}}, \bibinfo {author} {\bibfnamefont {S.-J.}\
			\bibnamefont {Xiong}}, \bibinfo {author} {\bibfnamefont {Q.}~\bibnamefont
			{Xu}}, \bibinfo {author} {\bibfnamefont {Z.}~\bibnamefont {Sun}}, \bibinfo
		{author} {\bibfnamefont {K.}~\bibnamefont {Chen}}, \bibinfo {author}
		{\bibfnamefont {F.}~\bibnamefont {Nori}},\ and\ \bibinfo {author}
		{\bibfnamefont {C.-P.}\ \bibnamefont {Yang}},\ }\bibfield  {title} {\bibinfo
		{title} {Experimental demonstration of quantum walks with initial
			superposition states},\ }\href {https://doi.org/10.1038/s41534-019-0155-x}
	{\bibfield  {journal} {\bibinfo  {journal} {npj Quantum Information}\
		}\textbf {\bibinfo {volume} {5}},\ \bibinfo {pages} {40} (\bibinfo {year}
		{2019})}\BibitemShut {NoStop}%
	\bibitem [{\citenamefont {Martinez-Dorantes}\ \emph {et~al.}(2018)\citenamefont
		{Martinez-Dorantes}, \citenamefont {Alt}, \citenamefont {Gallego},
		\citenamefont {Ghosh}, \citenamefont {Ratschbacher},\ and\ \citenamefont
		{Meschede}}]{PhysRevA.97.023410}%
	\BibitemOpen
	\bibfield  {author} {\bibinfo {author} {\bibfnamefont {M.}~\bibnamefont
			{Martinez-Dorantes}}, \bibinfo {author} {\bibfnamefont {W.}~\bibnamefont
			{Alt}}, \bibinfo {author} {\bibfnamefont {J.}~\bibnamefont {Gallego}},
		\bibinfo {author} {\bibfnamefont {S.}~\bibnamefont {Ghosh}}, \bibinfo
		{author} {\bibfnamefont {L.}~\bibnamefont {Ratschbacher}},\ and\ \bibinfo
		{author} {\bibfnamefont {D.}~\bibnamefont {Meschede}},\ }\bibfield  {title}
	{\bibinfo {title} {State-dependent fluorescence of neutral atoms in optical
			potentials},\ }\href {https://doi.org/10.1103/PhysRevA.97.023410} {\bibfield
		{journal} {\bibinfo  {journal} {Phys. Rev. A}\ }\textbf {\bibinfo {volume}
			{97}},\ \bibinfo {pages} {023410} (\bibinfo {year} {2018})}\BibitemShut
	{NoStop}%
	\bibitem [{\citenamefont {Sheludko}\ \emph {et~al.}(2008)\citenamefont
		{Sheludko}, \citenamefont {Bell}, \citenamefont {Anderson}, \citenamefont
		{Hofmann}, \citenamefont {Vredenbregt},\ and\ \citenamefont
		{Scholten}}]{PhysRevA.77.033401}%
	\BibitemOpen
	\bibfield  {author} {\bibinfo {author} {\bibfnamefont {D.~V.}\ \bibnamefont
			{Sheludko}}, \bibinfo {author} {\bibfnamefont {S.~C.}\ \bibnamefont {Bell}},
		\bibinfo {author} {\bibfnamefont {R.}~\bibnamefont {Anderson}}, \bibinfo
		{author} {\bibfnamefont {C.~S.}\ \bibnamefont {Hofmann}}, \bibinfo {author}
		{\bibfnamefont {E.~J.~D.}\ \bibnamefont {Vredenbregt}},\ and\ \bibinfo
		{author} {\bibfnamefont {R.~E.}\ \bibnamefont {Scholten}},\ }\bibfield
	{title} {\bibinfo {title} {State-selective imaging of cold atoms},\ }\href
	{https://doi.org/10.1103/PhysRevA.77.033401} {\bibfield  {journal} {\bibinfo
			{journal} {Phys. Rev. A}\ }\textbf {\bibinfo {volume} {77}},\ \bibinfo
		{pages} {033401} (\bibinfo {year} {2008})}\BibitemShut {NoStop}%
	\bibitem [{\citenamefont {Dadras}\ \emph {et~al.}(2018)\citenamefont {Dadras},
		\citenamefont {Gresch}, \citenamefont {Groiseau}, \citenamefont {Wimberger},\
		and\ \citenamefont {Summy}}]{PhysRevLett.121.070402}%
	\BibitemOpen
	\bibfield  {author} {\bibinfo {author} {\bibfnamefont {S.}~\bibnamefont
			{Dadras}}, \bibinfo {author} {\bibfnamefont {A.}~\bibnamefont {Gresch}},
		\bibinfo {author} {\bibfnamefont {C.}~\bibnamefont {Groiseau}}, \bibinfo
		{author} {\bibfnamefont {S.}~\bibnamefont {Wimberger}},\ and\ \bibinfo
		{author} {\bibfnamefont {G.~S.}\ \bibnamefont {Summy}},\ }\bibfield  {title}
	{\bibinfo {title} {Quantum walk in momentum space with a bose-einstein
			condensate},\ }\href {https://doi.org/10.1103/PhysRevLett.121.070402}
	{\bibfield  {journal} {\bibinfo  {journal} {Phys. Rev. Lett.}\ }\textbf
		{\bibinfo {volume} {121}},\ \bibinfo {pages} {070402} (\bibinfo {year}
		{2018})}\BibitemShut {NoStop}%
	\bibitem [{\citenamefont {Schreiber}\ \emph {et~al.}(2010)\citenamefont
		{Schreiber}, \citenamefont {Cassemiro}, \citenamefont
		{Poto\ifmmode~\check{c}\else \v{c}\fi{}ek}, \citenamefont {G\'abris},
		\citenamefont {Mosley}, \citenamefont {Andersson}, \citenamefont {Jex},\ and\
		\citenamefont {Silberhorn}}]{PhysRevLett.104.050502}%
	\BibitemOpen
	\bibfield  {author} {\bibinfo {author} {\bibfnamefont {A.}~\bibnamefont
			{Schreiber}}, \bibinfo {author} {\bibfnamefont {K.~N.}\ \bibnamefont
			{Cassemiro}}, \bibinfo {author} {\bibfnamefont {V.}~\bibnamefont
			{Poto\ifmmode~\check{c}\else \v{c}\fi{}ek}}, \bibinfo {author} {\bibfnamefont
			{A.}~\bibnamefont {G\'abris}}, \bibinfo {author} {\bibfnamefont {P.~J.}\
			\bibnamefont {Mosley}}, \bibinfo {author} {\bibfnamefont {E.}~\bibnamefont
			{Andersson}}, \bibinfo {author} {\bibfnamefont {I.}~\bibnamefont {Jex}},\
		and\ \bibinfo {author} {\bibfnamefont {C.}~\bibnamefont {Silberhorn}},\
	}\bibfield  {title} {\bibinfo {title} {Photons walking the line: A quantum
			walk with adjustable coin operations},\ }\href
	{https://doi.org/10.1103/PhysRevLett.104.050502} {\bibfield  {journal}
		{\bibinfo  {journal} {Phys. Rev. Lett.}\ }\textbf {\bibinfo {volume} {104}},\
		\bibinfo {pages} {050502} (\bibinfo {year} {2010})}\BibitemShut {NoStop}%
	\bibitem [{\citenamefont {Chandra}\ \emph {et~al.}(2023)\citenamefont
		{Chandra}, \citenamefont {Wu}, \citenamefont {Frank},\ and\ \citenamefont
		{Grieve}}]{IEEE.10141590}%
	\BibitemOpen
	\bibfield  {author} {\bibinfo {author} {\bibfnamefont {A.}~\bibnamefont
			{Chandra}}, \bibinfo {author} {\bibfnamefont {S.~J.}\ \bibnamefont {Wu}},
		\bibinfo {author} {\bibfnamefont {A.}~\bibnamefont {Frank}},\ and\ \bibinfo
		{author} {\bibfnamefont {J.~A.}\ \bibnamefont {Grieve}},\ }\bibfield  {title}
	{\bibinfo {title} {Compressive single-pixel read-out of single-photon quantum
			walks on a polymer photonic chip},\ }\href
	{https://doi.org/10.1109/JPHOT.2023.3281830} {\bibfield  {journal} {\bibinfo
			{journal} {IEEE Photonics Journal}\ }\textbf {\bibinfo {volume} {15}},\
		\bibinfo {pages} {1} (\bibinfo {year} {2023})}\BibitemShut {NoStop}%
	\bibitem [{\citenamefont {Hines}\ and\ \citenamefont
		{Stamp}(2007)}]{PhysRevA.75.062321}%
	\BibitemOpen
	\bibfield  {author} {\bibinfo {author} {\bibfnamefont {A.~P.}\ \bibnamefont
			{Hines}}\ and\ \bibinfo {author} {\bibfnamefont {P.~C.~E.}\ \bibnamefont
			{Stamp}},\ }\bibfield  {title} {\bibinfo {title} {Quantum walks, quantum
			gates, and quantum computers},\ }\href
	{https://doi.org/10.1103/PhysRevA.75.062321} {\bibfield  {journal} {\bibinfo
			{journal} {Phys. Rev. A}\ }\textbf {\bibinfo {volume} {75}},\ \bibinfo
		{pages} {062321} (\bibinfo {year} {2007})}\BibitemShut {NoStop}%
	\bibitem [{\citenamefont {Balu}\ \emph {et~al.}(2018)\citenamefont {Balu},
		\citenamefont {Castillo},\ and\ \citenamefont {Siopsis}}]{Balu_2018}%
	\BibitemOpen
	\bibfield  {author} {\bibinfo {author} {\bibfnamefont {R.}~\bibnamefont
			{Balu}}, \bibinfo {author} {\bibfnamefont {D.}~\bibnamefont {Castillo}},\
		and\ \bibinfo {author} {\bibfnamefont {G.}~\bibnamefont {Siopsis}},\
	}\bibfield  {title} {\bibinfo {title} {Physical realization of topological
			quantum walks on ibm-q and beyond},\ }\href
	{https://doi.org/10.1088/2058-9565/aab823} {\bibfield  {journal} {\bibinfo
			{journal} {Quantum Science and Technology}\ }\textbf {\bibinfo {volume}
			{3}},\ \bibinfo {pages} {035001} (\bibinfo {year} {2018})}\BibitemShut
	{NoStop}%
	\bibitem [{\citenamefont {Wadhia}\ \emph {et~al.}(2024)\citenamefont {Wadhia},
		\citenamefont {Chancellor},\ and\ \citenamefont {Kendon}}]{Wadhia2024}%
	\BibitemOpen
	\bibfield  {author} {\bibinfo {author} {\bibfnamefont {V.}~\bibnamefont
			{Wadhia}}, \bibinfo {author} {\bibfnamefont {N.}~\bibnamefont {Chancellor}},\
		and\ \bibinfo {author} {\bibfnamefont {V.}~\bibnamefont {Kendon}},\
	}\bibfield  {title} {\bibinfo {title} {Cycle discrete-time quantum walks on a
			noisy quantum computer},\ }\href
	{https://doi.org/10.1140/epjd/s10053-023-00795-2} {\bibfield  {journal}
		{\bibinfo  {journal} {Eur. Phys. J. D}\ }\textbf {\bibinfo {volume} {78}},\
		\bibinfo {pages} {29} (\bibinfo {year} {2024})}\BibitemShut {NoStop}%
	\bibitem [{\citenamefont {Nandi}\ \emph {et~al.}(2024)\citenamefont {Nandi},
		\citenamefont {Singha}, \citenamefont {Datta}, \citenamefont {Saha},\ and\
		\citenamefont {Chakrabarti}}]{Nandi24}%
	\BibitemOpen
	\bibfield  {author} {\bibinfo {author} {\bibfnamefont {B.}~\bibnamefont
			{Nandi}}, \bibinfo {author} {\bibfnamefont {S.}~\bibnamefont {Singha}},
		\bibinfo {author} {\bibfnamefont {A.}~\bibnamefont {Datta}}, \bibinfo
		{author} {\bibfnamefont {A.}~\bibnamefont {Saha}},\ and\ \bibinfo {author}
		{\bibfnamefont {A.}~\bibnamefont {Chakrabarti}},\ }\href@noop {} {\bibinfo
		{title} {Robust implementation of discrete-time quantum walks in any
			finite-dimensional quantum system}} (\bibinfo {year} {2024}),\ \bibinfo
	{note} {arXiv:2408.00530}\BibitemShut {NoStop}%
	\bibitem [{\citenamefont {Preskill}(2018)}]{Preskill2018quantumcomputingin}%
	\BibitemOpen
	\bibfield  {author} {\bibinfo {author} {\bibfnamefont {J.}~\bibnamefont
			{Preskill}},\ }\bibfield  {title} {\bibinfo {title} {Quantum {C}omputing in
			the {NISQ} era and beyond},\ }\href
	{https://doi.org/10.22331/q-2018-08-06-79} {\bibfield  {journal} {\bibinfo
			{journal} {{Quantum}}\ }\textbf {\bibinfo {volume} {2}},\ \bibinfo {pages}
		{79} (\bibinfo {year} {2018})}\BibitemShut {NoStop}%
	\bibitem [{\citenamefont {Bharti}\ \emph {et~al.}(2022)\citenamefont {Bharti},
		\citenamefont {Cervera-Lierta}, \citenamefont {Kyaw}, \citenamefont {Haug},
		\citenamefont {Alperin-Lea}, \citenamefont {Anand}, \citenamefont {Degroote},
		\citenamefont {Heimonen}, \citenamefont {Kottmann}, \citenamefont {Menke},
		\citenamefont {Mok}, \citenamefont {Sim}, \citenamefont {Kwek},\ and\
		\citenamefont {Aspuru-Guzik}}]{RevModPhys.94.015004}%
	\BibitemOpen
	\bibfield  {author} {\bibinfo {author} {\bibfnamefont {K.}~\bibnamefont
			{Bharti}}, \bibinfo {author} {\bibfnamefont {A.}~\bibnamefont
			{Cervera-Lierta}}, \bibinfo {author} {\bibfnamefont {T.~H.}\ \bibnamefont
			{Kyaw}}, \bibinfo {author} {\bibfnamefont {T.}~\bibnamefont {Haug}}, \bibinfo
		{author} {\bibfnamefont {S.}~\bibnamefont {Alperin-Lea}}, \bibinfo {author}
		{\bibfnamefont {A.}~\bibnamefont {Anand}}, \bibinfo {author} {\bibfnamefont
			{M.}~\bibnamefont {Degroote}}, \bibinfo {author} {\bibfnamefont
			{H.}~\bibnamefont {Heimonen}}, \bibinfo {author} {\bibfnamefont {J.~S.}\
			\bibnamefont {Kottmann}}, \bibinfo {author} {\bibfnamefont {T.}~\bibnamefont
			{Menke}}, \bibinfo {author} {\bibfnamefont {W.-K.}\ \bibnamefont {Mok}},
		\bibinfo {author} {\bibfnamefont {S.}~\bibnamefont {Sim}}, \bibinfo {author}
		{\bibfnamefont {L.-C.}\ \bibnamefont {Kwek}},\ and\ \bibinfo {author}
		{\bibfnamefont {A.}~\bibnamefont {Aspuru-Guzik}},\ }\bibfield  {title}
	{\bibinfo {title} {Noisy intermediate-scale quantum algorithms},\ }\href
	{https://doi.org/10.1103/RevModPhys.94.015004} {\bibfield  {journal}
		{\bibinfo  {journal} {Rev. Mod. Phys.}\ }\textbf {\bibinfo {volume} {94}},\
		\bibinfo {pages} {015004} (\bibinfo {year} {2022})}\BibitemShut {NoStop}%
	\bibitem [{\citenamefont {AbuGhanem}\ and\ \citenamefont
		{Eleuch}(2024)}]{AbuGhanem24}%
	\BibitemOpen
	\bibfield  {author} {\bibinfo {author} {\bibfnamefont {M.}~\bibnamefont
			{AbuGhanem}}\ and\ \bibinfo {author} {\bibfnamefont {H.}~\bibnamefont
			{Eleuch}},\ }\bibfield  {title} {\bibinfo {title} {Nisq computers: A path to
			quantum supremacy},\ }\href {https://doi.org/10.1109/ACCESS.2024.3432330}
	{\bibfield  {journal} {\bibinfo  {journal} {IEEE Access}\ }\textbf {\bibinfo
			{volume} {12}},\ \bibinfo {pages} {102941} (\bibinfo {year}
		{2024})}\BibitemShut {NoStop}%
\end{thebibliography}

\end{document}